\documentclass[prd,superscriptaddress,showpacs,twocolumns]{revtex4}
\usepackage{amsmath}
\usepackage{graphicx}
\usepackage{dsfont}
\usepackage{subfigure}
\def \D {\tilde{\nabla}}
\def \curl {\mbox{curl}\,}
\def \ep {\varepsilon}
\def\l{\label}

\def\Th{\Theta}
\def\dd{\mathcal D}

\def\sig{\sigma}
\def\om{\omega}
\def\udot{\dot{u}}
\def\nab{\nabla}
\def\3nab{\tilde{\nabla}}

\def\lgl{\langle}
\def\rgl{\rangle}

\def\nn{\nonumber}
\def\c{\mbox{curl}}

\def\hsp5{\hspace{5mm}}

\def\case#1/#2{\textstyle\frac{#1}{#2}}

\def\be {\begin{equation}}
\def\ee {\end{equation}}
\def\ber {\begin{eqnarray}}
\def\eer {\end{eqnarray}}
\def\bea {\begin{eqnarray}}
\def\eea {\end{eqnarray}}

\def\bc {\begin{center}}
\def\ec {\end{center}}
\def\case#1/#2{\frac{#1}{#2}}
\def\rf#1{(\ref{#1})}

\def\cqg{{\it Class. Quantum Grav.}\ }

\def\etal\;{{\it et al.}}

\begin{document}

\title{Unifying the study of background dynamics and perturbations in $f(R)$-gravity}

\author{Sante Carloni}
\affiliation{Institut de Cincies de l'Espai (CSIC-IEEC)
Campus UAB -  Facultat de Ciencies
Torre C5  Parell,  2da Planta
E-08193 Bellaterra  (Barcelona)
Spain }
 \author{Kishore N. Ananda}
 \affiliation{ Department of Mathematics and Applied\ Mathematics,
University of Cape Town, South Africa.}
 \author{Peter K. S. Dunsby}
\affiliation{ Department of Mathematics and Applied\ Mathematics,
University of Cape Town, South Africa.}
\affiliation{\ South African
Astronomical Observatory, Observatory Cape Town, South Africa.}
 \author{Mohamed E. S. Abdelwahab}
 \affiliation{ Department of Mathematics and Applied\ Mathematics,
University of Cape Town, South Africa.}

\begin{abstract}
In this paper we show how the covariant gauge invariant equations for the evolution of scalar, vector and tensor perturbations for a generic $f(R)$-gravity theory can be recast in order to exploit the power of dynamical system methodology. In this way, recent results describing the dynamics of the background FRW model can be easily combined with these equations to reveal important details pertaining to the evolution of cosmological models in fourth order gravity.
\end{abstract}

\date{\today}
 \maketitle

\section{Introduction}
The observational evidence for the accelerated expansion rate of the universe, and the introduction of the concept of Dark Energy has put theoretical cosmology into crisis. This is due to the fact that despite an increasing amount and quality of data, no model has been proposed thus far that is able to give a completely satisfactory theoretical explanation of these observations.

Among the many different ways to achieve cosmic acceleration, the modification of gravity and in particular higher order gravity has recently gained much attention \cite{revnostra,Odintsov, Carroll,star2007,Sotiriou:2008rp}. The reason for this popularity is due to the fact that these models provide a somewhat more natural explanation of the cosmic acceleration: this effect is due to corrections to Einstein gravity which are directly related to the characteristic properties of the gravitational interaction. Most investigations of higher order gravity have focused on Fourth Order Gravity (FOG), i.e., on gravitational Lagrangian's in which the corrections are at most of order four in the metric and in what follows we will also focus on these models.

Because the field equations resulting from FOG are extremely complicated,  difficult conceptual and technical issues arise which need to be resolved in order to uncover the detailed physics of these models. Consequently it is important  to develop new methods which are able to assist in resolving these problems. Two such approaches, the {\it dynamical systems approach to cosmology} and the {\it covariant approach to cosmological perturbations}, have proved particularly useful in this respect.

The dynamical system approach, first developed by Collins \cite{collins} and extensively reviewed in the book by Ellis and Wainwright  \cite{ellisbook}  has proved to be an important tool in the understanding of cosmology of $f(R)$-gravity models. In fact, studying cosmological models using these techniques has the advantage of providing a relatively simple method for obtaining exact solutions, which appear as fixed points of the system and obtaining a global picture of the dynamics of these models. Consequently, such an analysis allows for an efficient preliminary investigation of these theories, allowing one to identify specific models which merit further investigation. In a series of recent  papers, a  wide range of features of  $f(R)$ cosmology have been presented, ranging from an analysis of the standard Friedmann-Lema\^{\i}tre-Robertson-Walker (FLRW) models \cite{Amendola,GenDynSys,RnGrav} to a discussion of the properties of the Einstein universe and the isotropization of Bianchi models \cite{Leach:2006br, Barrow:2006xb,Goheer:2007wu}.

The 1+3 covariant approach was developed originally to study the evolution of linear perturbations of FRW models in General Relativity in\cite{EllisCovariant,EB,EBH,BDE,DBE,BED,DBBE,conserved} with great success. For our purposes this approach has two major advantages. Firstly,  a specific recasting of the field equations allows one to easily make extensions to a wide range of modified gravity theories including Braneworlds \cite{brane} and FOG \cite{SantePertSca,StructForm}. Secondly the structure of the formalism is such that,  unlike other approaches , it is possible to keep track of the physical meaning of the equations at any stage of the calculations, which can be crucial when one modifies the theory of gravity.

In particular, in a number of recent papers \cite{SantePertSca,StructForm,PRL} we derived the evolution equations for scalar and tensor perturbations of a subclass of fourth order theories of gravity characterized by an action which is a general analytic function of the Ricci scalar. The results obtained in \cite{SantePertSca,StructForm,PRL} clearly demonstrated that the evolution of scalar perturbations is determined by a fourth order differential equation rather than a second order one. This implies that the evolution of the density fluctuations contains, in general, four modes rather than two and can give rise to a more complex evolution than what is obtained in General Relativity. It was also found that the perturbations depend on the scale for any value of the equation of state parameter of standard matter (while in GR the evolution of the dust perturbations are not scale dependent) and that there is a characteristic scale-dependent signature in the matter power spectrum \cite{PRL}. This means that in FOG the evolution of super-horizon and sub-horizon  perturbations are different. It also turns out that the growth of large density fluctuations can occur also in backgrounds in which the expansion rate is increasing in time. This surprising result is strikingly different with what one finds in GR and could provide a strong constraint on some FOG theories using the ISW effect. In addition to that the structure of the general fourth order perturbation equations and the analysis of scalar perturbations lead to the discovery of a characteristic signature of fourth order gravity in the matter power spectrum \cite{StructForm}, the details of which have not seen before in other works in this area. This could provide a crucial test for fourth order gravity on cosmological scales.

The aim of this paper is to combine dynamical systems methods with linear perturbation theory in such a manner that one is able to apply directly the results coming from the former to the latter and is able to gain a semi-qualitative idea of both the behavior of the background and that of the first order perturbations in a general $f(R)$-gravity theory. In order to achieve this we will express the coefficients of the perturbation equations in terms of the dynamical system variables in such a way that at any fixed point it will correspond a set of perturbation equations and as a consequence an evolution law for the linear fluctuations.

The paper is divided as follow. In Section 2 we give some basic equations. In Section 3 we present a formalism that allows one to investigate a large class of  $f(R)$-gravity models. In Section 4 we present the equations for the evolution of the scalar, vector and tensor perturbations in a generic $f(R)$-gravity theory written in terms of the dynamical system variables. In Section 5 we consider two simple examples. Finally, Section 6 is devoted to the conclusions.

Unless otherwise specified, natural units ($\hbar=c=k_{B}=8\pi G=1$)
will be used throughout this paper, Latin indices run from 0 to 3.
The symbol $\nabla$ represents the usual covariant derivative and
$\partial$ corresponds to partial differentiation. We use the
$-,+,+,+$ signature and the Riemann tensor is defined by
\begin{equation}
R^{a}{}_{bcd}=W^a{}_{bd,c}-W^a{}_{bc,d}+ W^e{}_{bd}W^a{}_{ce}-
W^f{}_{bc}W^a{}_{df}\;,
\end{equation}
where the $W^a{}_{bd}$ are the Christoffel symbols (i.e. symmetric in
the lower indices), defined by
\begin{equation}
W^a_{bd}=\frac{1}{2}g^{ae}
\left(g_{be,d}+g_{ed,b}-g_{bd,e}\right)\;.
\end{equation}
The Ricci tensor is obtained by contracting the {\em first} and the
{\em third} indices
\begin{equation}\label{Ricci}
R_{ab}=g^{cd}R_{acbd}\;.
\end{equation}
Finally the Hilbert--Einstein action in the presence of matter is
given by
\begin{equation}
{\cal A}=\int d x^{4} \sqrt{-g}\left[\frac{1}{2}R+ L_{m}\right]\;.
\end{equation}

\section{Basic equations}
In four dimensional homogeneous and isotropic spacetimes i.e.,  Friedmann Lema\^{\i}tre Robertson Walker (FLRW) universes,  the most general action for fourth order gravity can be written as an analytic function of the Ricci scalar only:
\begin{equation}\label{lagr f(R)}
\mathcal{A}=\int d^4 x \sqrt{-g}\left[ f(R)+{\cal L}_{m}\right]\;,
\end{equation}
where $\mathcal{L}_m$ represents the matter contribution.
Varying the action with respect to the metric gives the generalization of the Einstein equations:
\begin{equation}\label{eq:einstScTn}
f'G_{ab}=f'\left(R_{ab}-\frac{1}{2}\,g_{ab} R\right)=T
_{ab}^{m}+\frac{1}{2}g_{ab} \left(f-R f'\right) +\nab_b\nab_a f'-
g_{ab}\nab_c\nab^c f'\;,
\end{equation}
where $f=f(R)$, $f'= \displaystyle{\frac{d f(R)}{dR}}$, and
$\displaystyle{T^{M}_{ab}=\frac{2}{\sqrt{-g}}\frac{\delta
(\sqrt{-g}\mathcal{L}_{m})}{\delta g_{ab}}}$ represents the
stress energy tensor of standard matter. These equations reduce to
the standard Einstein field equations when $f(R)=R$. It is crucial
for our purposes to be able to write \rf{eq:einstScTn} in the form
\begin{equation}
\label{eq:einstScTneff}
 G_{ab}=\tilde{T}_{ab}^{m}+T^{R}_{ab}=T^{tot}_{ab}\,,
 \end{equation}
where $\displaystyle{\tilde{T}_{ab}^{m}=\frac{ T_{ab}^{m}}{f'}}$ and
\begin{eqnarray}\label{eq:TenergymomentuEff}
T_{ab}^{R}=\frac{1}{f'}\left[\frac{1}{2}g_{ab} \left(f-R f'\right)
+\nab_b\nab_a f'- g_{ab}\nab_c\nab^cf'\right], \label{eq:semt}
\end{eqnarray}
represent two effective ``fluids": the  {\em curvature ``fluid"}
(associated with $T^{R}_{ab}$) and  the {\em effective matter
``fluid"} (associated with $\tilde{T}_{ab}^{m}$) \cite{revnostra,SantePertSca}. This step is
important because it allows us to treat fourth order gravity as
standard Einstein gravity plus two ``effective" fluids.  The details of the conservation properties of these effective fluids have been given in \cite{SantePertSca}. In particular, it has been shown that, no matter how the  effective fluids behave, standard matter still follows the usual conservation equations $T_{ab}^{m\ ;b}=0$.

\section{Dynamical System Approach and f(R) gravity}
The DS approach \cite{ellisbook} has been used with great success in the analysis of the background evolution of cosmological models with fourth order gravity. In \cite{Amendola, GenDynSys} it was shown that using the dimensionless variables:
\begin{eqnarray}\label{dynsysvar}
x = \frac{3\dot{f'}}{f' \Theta}, \qquad y = \frac{3 R}{2 \Theta^2},  \qquad z = \frac{3f}{2 f' \Theta^2}, \qquad
\Omega = \frac{3 \mu_m}{ f' \Theta^2},  \qquad \mathds{\chi} =\frac{9 K}{S^2 \Theta^2}\;,
\end{eqnarray}
together with the characteristic variable
\begin{equation}\label{q}
\mathds{Q}\,\equiv\displaystyle\left(\frac{d \log{f'}}{d\log{R}}\right)^{-1}\,=\,\frac{f'}{Rf''}\,
\end{equation}
and logarithmic time  $N=|\ln S|$,  the cosmological equations are equivalent to the autonomous system:
\begin{eqnarray}\label{DynsysNoX}
\frac{dy}{d N} &=&\varepsilon  y [2 \chi-2 y+\mathds{Q}(-\chi+y-z+\Omega-1)+4]\,, \nonumber \\
\frac{dz}{d N} &=&\varepsilon z (3 \chi-3y+z-\Omega+5)+\mathds{Q} \epsilon y(-\chi+y-z+\Omega-1)\,,  \\
\frac{d\Omega}{d N} &=& -\varepsilon\Omega (3 w-3 \chi+3y-z+\Omega-2)\,, \nonumber \\
\frac{d\chi}{d N} &=&2 \varepsilon  \chi(\chi-y+1)\,, \nonumber \\
1 &= &  y + \Omega - x- z - \chi \;. \nonumber
\end{eqnarray}
where $\varepsilon=|\Theta|/\Theta$ \footnote{We have chosen here to use the constraint to eliminate the variable $x$. This is different to what has been done in the other works on this subject. The reason for this choice is due to the fact that, since we will eventually express the coefficients of the perturbation equations in terms of the dynamical system variables, it is more useful to retain the variables which have a more direct physical meaning.}. This system allows one to analyze many interesting fourth order gravity cosmological models and leads to the result that some of them present cosmic histories which posses a transient Friedmann phase and evolve naturally towards an accelerated (DE-like) expansion phase.

A detailed analysis of the properties and caveats of this method can be found in \cite{Amendola, GenDynSys}, here it is important only to remind the reader that the solutions associated with the fixed points can be found by substituting the coordinates of the fixed points into the system
\begin{eqnarray}\label{solsystem}
\dot{\Theta} &=&\gamma \Theta^{2}\;, \qquad \gamma=\frac{-2 + y_i - \chi_i}{3} \,,\label{solsystem1}\\
\dot{\mu}_m &=& -\frac{(1+w)}{\gamma\; t}\mu_m \label{solsystem2}\,,
\end{eqnarray}
where the subscript ``$i$" stands for the value of a generic quantity at a fixed point.
This means that  for $\gamma\neq 0$, the general solutions  can be written as
\begin{eqnarray}\label{solsystemalp}
a &=&a_{0}(t-t_{0})^{1/3\gamma }\;,  \\
\mu_m &=& a_{0}(t-t_{0})^{-\frac{(1+w)}{3\gamma}} \,.
\end{eqnarray}
The  expression above gives the solution for the scale factor and the evolution
of the energy density for every fixed point in which
$\gamma\neq 0$. When $\gamma=0$ the \rf{solsystem1} reduces to $\dot{\Theta}=0$ which correspond to either a static or a de Sitter solution.

It is important to stress at this point that sometimes the solutions associated with the fixed points are ``non physical" i.e., they do not satisfy the cosmological equations. Then one might ask how is it possible that, although derived from the cosmological equations themselves, our dynamical system equations give non physical solutions. The reason for this apparent contradiction needs to be looked for in the very structure of the dynamical system approach. As we have seen, the condition to obtain the fixed points of the system (\ref{DynsysNoX}) --as well as every dynamical system-- is to set the first derivative of the dynamical variables to zero (i.e., to set the left hand side of  (\ref{DynsysNoX}) to zero)  and solve the system obtained. Such a step, in the standard dynamical system theory is usually trivial, however, in our specific formulation this step becomes much more subtle. In fact,  the requirement of the existence of a fixed point is also imposes the requirement that all the variables acquire a constant value (or equivalently that their first derivative with respect to the time coordinate is zero). This is equivalent to an additional system of equations that is not necessarily satisfied by the solutions of the system. For example, in
\begin{equation}
\mathbf{x'}=\mathbf{F}(\mathbf{x})\;,
\end{equation}
the condition to obtain the fixed points would be
\begin{equation}\label{FPcond}
\mathbf{F}(\mathbf{x})=0\qquad\mbox{but also}\qquad \mathbf{x}=const \;\;\;(\mathbf{x'}=0)\,,
\end{equation}
as mentioned earlier, the second system is usually trivially satisfied. In the formalism above, however, these variables are a combination of quantities appearing in the cosmological equations. This means that $\mathbf{x}=const$ becomes a set of conditions to be satisfied by all the physical fixed points of the system. In GR, because of the structure of the variables \cite{ellisbook,Goliath:1998na}, these constraints are  automatically satisfied.  In fact, in this case one has $\Omega_{GR}=3\mu^{m} / \Theta^2$  and $\chi_{GR}=9 K / S^2 \Theta^2$ which means
 \begin{eqnarray}
&&\frac{d\Omega_{GR}}{d N}= 0 \qquad\Rightarrow\qquad\Omega_{GR}=const. \qquad\Rightarrow\qquad  \mu^{m}\propto\Theta^2\,, \\&& \frac{d\chi_{GR}}{d N}= 0 \qquad\Rightarrow\qquad\chi_{GR}=const. \qquad\Rightarrow\qquad S^2\propto K \Theta^2\,,
\end{eqnarray}
so the physical points can either have  $K=0$ and $\mu^{m}\propto \Theta^2$ or $K\neq0$, $S\propto \Theta$. Both the fixed point that one obtains  (corresponding to Milne and Friedmann solutions) satisfy these criteria.

This then raises the following question:  how do we consider the non physical fixed points? The answer depends on the stability. If the fixed point is unstable then, although the solution associated with the fixed point does not satisfy the cosmological equations, it can be used to approximate the behavior in neighborhood of the fixed point. Orbits nearby this point will bounce on or run away from it, but they will never reach  it. Instead in the case in which the point is stable, the set of orbits approaching the point will not correspond to any physical evolution for the system and the dynamical system approach fails to give an appropriate description of the cosmological evolution. This imposes a intrinsic limitation in predictive power of this approach and has to be taken into account to avoid incorrect conclusions.

Another consequence of this limitation is that the structure of the variables characterize the type of solution associated with the fixed points and, consequently, the fixed points one derives with a dynamical system formalism are not necessarily the complete set of the elementary solutions of the system. This is also important because it implies that the absence of a specific cosmic history in a dynamical systems formalism does not necessarily indicate that this cosmic history cannot be realized, but only that the specific formalism used is not able to show its presence.

\section{Covariant approach to perturbation theory}
The form \rf{eq:einstScTneff} of the field equations allows us to use directly the covariant gauge invariant approach  \cite{EllisCovariant,EB,EBH,DBE,BED,BDE,DBBE,conserved}  in the same way as presented in \cite{SantePertSca,StructForm}.
Following these references we will choose a frame  $u^{m}_a$ comoving with standard matter ({\it the matter energy frame}) \footnote{Obviously this also means that the equations which we present will change, in general, if a different choice of frame is made. This will not happen to the equations for the tensor perturbations because in that case the equations are frame invariant to linear order}. We will also assume that in $u^{m}_a$, the standard matter is a barotropic perfect fluid with an equation of state $p_m=w \mu_m$.

Once the frame has been chosen, the derivation of the kinematical quantities can be obtained in the standard manner \cite{EllisCovariant}. In particular the derivative along the matter fluid flow lines is defined by $\dot{X}=u_a\nabla^aX$ and the projected covariant derivative operator orthogonal to $u^a$ is given by $\3nab_a=h^b{}_a\nabla_b$. With these definitions we can define the key kinematic quantities of the cosmological model: the expansion $\Theta$, the shear $\sigma_{ab}$, the vorticity $\omega_{ab}$ and the acceleration  $a_a$. The general propagation equations for these kinematic variables in any spacetime correspond to the so called {\em 1+3 covariant equations} \cite{EllisCovariant} and are given for completeness in Appendix \ref{CovID}.

The definition of a frame $u^a$ also allows us to obtain an irreducible
decomposition of the stress energy momentum tensor. In a general
frame and for a general tensor $T_{{a}{{b}}}$ one obtains:
\begin{equation}\label{Tdecomp}
T_{{a}{{b}}}=\mu u_a
u_{{b}}+ph_{{a}{{b}}}+2q_{(a}u_{{{b}})}+\pi_{{{a}}{{b}}}\,,
\end{equation}
where $\mu$ and $p$ are the energy density and isotropic pressure,
$q_{{{a}}}$ is the energy flux and $\pi_{{{a}}{{b}}}$
is the anisotropic pressure.

In this way, relative to $u^a_m$, $T^{tot}_{ab}$ can be decomposed as
\begin{eqnarray}\label{mupitot}
\mu^{\rm tot}\,&=&T^{\rm tot}_{ab}u^{a}u^{b}\,=\,\tilde{\mu}^{\,
m}+\mu^{\,R}\,,\qquad
p^{\rm tot}\,=\frac{1}{3}T^{\rm tot}_{ab}h^{ab}\,=\,\tilde{p}^{\, m}+p^{\,R}\;,
\end{eqnarray}
\begin{eqnarray}\label{qpaitot}
q^{\rm tot}_{a}\,&=&-T^{\rm
tot}_{bc}h_{a}^{b}u^{c}\,=\,\tilde{q}^{\,
m}_{a}+q^{\,R}_{a}\,,\qquad \pi^{\rm tot}_{ab}\,=\,T^{\rm
tot}_{cd}h_{<a}^{c}h_{b>}^{d}\,=\,\tilde{\pi}^{\,
m}_{ab}+\pi^{\,R}_{ab}\,,
\end{eqnarray}
with
\begin{eqnarray}
\tilde{\mu}^{\,m}\,&=&\,\frac{\mu^{\,m}}{f'}\,,\qquad
\tilde{p}^{\,m}\,=\,\frac{p^{\,m}}{f'}\,,\qquad
\tilde{q}^{\,m}_{a}\,=\,\frac{q^{\,m}_{a}}{f'}\,,\qquad
\tilde{\pi}^{\,m}_{ab}=\,\frac{\pi^{\,m}_{ab}}{f'}\, .
\end{eqnarray}
Since we assume that in our frame standard matter is a perfect fluid,
$q^{\,m}_{a}$ and $\pi^{\,m}_{ab}$ are zero, so that the last two
quantities above also vanish.

The effective thermodynamical quantities for the curvature ``fluid"
are
\begin{eqnarray}\label{thermoHO}
&&\mu^{R}\,=\,\frac{1}{f'}\left[\frac{1}{2}(R f'-f)-\Theta
f''\dot{R}+f''\tilde{\nabla}^2{R}+f'''\D^{a}{R}\D_{a}{R}\right]\;,\\
&&p^{R}\,=\,\frac{1}{f'}\left[\frac{1}{2}(f-R
f')+f''\ddot{R}+f'''\dot{R}^2+\frac{2}{3}\Theta
f''\dot{R}-\frac{2}{3}f''\tilde{\nabla}^2{R} +\right. \nonumber\\
&& \qquad\left.  -\frac{2}{3}f'''\D^{a}{R}\D_{a}{R}+f''
\,a_b\D^b{R}\right]\;,\\
&&q^{R}_a\,=\,-\frac{1}{f'}\left[f'''\dot{R}\D_{a}R+f''\D_{a}\dot{R}-\frac{1}{3}\Theta f''
\D_{a}R\right]\;,\\
&&\pi^{R}_{ab}\,=\,\frac{1}{f'}\left[f''\D_{\lgl
a}\D_{b\rgl}R+f'''\D_{\lgl a}{R}\D_{b\rgl}{R}-\sigma_{a
b}\dot{R}\right]\,.\label{piR}
\end{eqnarray}
The twice contracted Bianchi Identities lead to evolution equations for
$\mu^{\,m}$, $\mu^{R}$, $q^{R}_a$ and are given in Appendix  \ref{CovID}.

Using the quantities defined above, and the equations given in Appendix  \ref{CovID}, we are able to write both the evolution equations for the background and  ones of the first order perturbations. As in \cite{SantePertSca,StructForm}  we will consider a background that is homogeneous and isotropic, i.e., a FLRW model. In this background the cosmological equations for a generic $f(R)$ read:
\begin{eqnarray}\label{f}
&&\Theta^2\,=\,3\tilde{\mu}^{m} + 3\mu^{R}-\frac{3\tilde{R}}{2}\;,\\
&&\dot{\Theta}+{\textstyle\frac{1}{3}}\Theta^2
+{\textstyle\frac{1}{2}}(\tilde{\mu}^{m} + 3\tilde{p}^{m})
        +{\textstyle\frac{1}{2}}({\mu}^{R} + 3{p}^{R})=0\;,\\
&& \dot{\mu}^m\,+ \,\Theta\,(\mu^m+{p^m})=0\;,
\end{eqnarray}
where $\tilde{R}=6K/S^2$ is the 3-Ricci scalar, $K=0,\pm1$ and $S$ is the scale factor.

\section{Perturbation Equations and DSA}
In this section we will rewrite the evolution equations for scalar, vector and tensor perturbations in terms of the dynamical system variables. As we will see, the requirement to obtain a closed form for the coefficients will require a redefinition of some of the variables in the equations.

Once this has been done the behavior of the perturbations at a fixed point of the system  \rf{DynsysNoX} can be inferred in all generality. This can be extremely useful in determining  the set of cosmic histories which are compatible with observations for a given $f(R)$ model.

\subsection{Scalar perturbations}
The evolution equations for the scalar perturbations for a generic $f(R)$ theory of gravity in the covariant approach were derived in \cite{Li:2008ai, SantePertSca}  and successively  analyzed in detail in \cite{StructForm}. These types of perturbations are characterized by the quantities
\begin{equation}\label{ScaVar}
\Delta^{m}=\frac{S^2}{\mu^{m}}\3nab^2\mu^{m}\,,\qquad
Z=S^2\3nab^2\Theta\,,\qquad C=S^{4}\3nab^2\tilde{R}\,,\qquad{\cal R}=S^2\3nab^2 R\,,\qquad\Re=S^2\D \dot{R}\;,
\end{equation}
and their evolution equations are given in Appendix \ref{AppB}. In order to express these equations in terms of the dynamical system variables we first have to convert them into equations in which the independent variable is $N$. In addition to that one can only obtain a closed form for the coefficients if  the curvature variable is redefined as
\begin{equation}
\mathds{R}=S^2\3nab^2 \ln [f'(R)]\,.
\end{equation}
Substituting for the new curvature variable and using the definitions \rf{dynsysvar}, the perturbation equations developed in harmonics can be written as
\begin{eqnarray} \label{EqScaIIOrdDynSys}
   && \frac{d^{2} \,\Delta^{(\ell)}}{d N^{2}}+\mathcal{A}_1\; \frac{d \Delta^{(\ell)}}{d N}+\mathcal{B}_1\;\Delta^{(\ell)}+
  \mathcal{C} _1\;\mathds{R}^{(\ell)} +\mathcal{D}_1\;\frac{d \mathds{R}^{(\ell)}}{d N}=0\,,\\&&
 \frac{d^{2} \mathds{R}^{(\ell)}}{d N^{2}}+\mathcal{E}_1\,\frac{d \mathds{R}^{(\ell)}}{d N}+\mathcal{F}_1\;\mathds{R}^{(\ell)}+ \mathcal{G}_1 \; \Delta^{(\ell)}+\mathcal{H}_1\,\frac{d \Delta^{(\ell)}}{d N}=0\,.
\end{eqnarray}
where
\begin{eqnarray}
&&\mathcal{A}_1=\varepsilon(1-3 w  +z -\Omega )\,, \\
&&\mathcal{B}_1=-3(2 w z-3 w\mathcal{K}+ (1-w) \Omega)\,,\\
&&\mathcal{C}_1=-3 (w+1) (z-\mathds{Q} y-3 \mathcal{K}-\Omega)\,,\\
&&\mathcal{D}_1=3 \varepsilon (w+1)\,,
\\
&&\mathcal{E}_1=-\varepsilon  (3 \chi-3 y+2 z-2 \Omega+1)\,,\\
&&\mathcal{F}_1=4 z-2 \mathds{Q} y-9\mathcal{K}+(3 w-1) \Omega\,,\\
&&\mathcal{G}_1=\frac{w(4  y-8  z)-\left(3 w^2-4 w+1\right) \Omega }{w+1}\,,\\
 &&\mathcal{H}_1=-\frac{\varepsilon(w-1)}{w+1} (\chi-y+z-\Omega +1)\,,
 \end{eqnarray}
and the form of  $\mathcal{K}(N)$ is given by
\begin{equation}
\frac{d \mathcal{K}}{d N}=2 \varepsilon  \mathcal{K} (\chi-y+1)\,.
\end{equation}
Note that in this form the above equations are such that two different forms of the Lagrangian have the same evolution of the scalar perturbations if they both have a fixed point with the same coordinates {\it and} and in this fixed point $\mathds{Q}$ is the same. As we will see this will allow us to attach a fixed point to a certain evolution law of the perturbations.

\subsection{Vector Perturbations}
In order to  analyze the evolution of the vector perturbations one has to extract the vector parts of the variables in the 1+3 equations (\ref{1+3eqRayHO}-\ref{eq:cons3}) and the propagation equations for $D_a,Z_a, \mathcal{R}_a$ and $\Re_a$ in Appendix \ref{AppB}. In our specific case some important facts have to be noted. First,  looking at \rf{thermoHO} the heat flux $q_a$ and the anisotropic pressure $\pi_{ab}$ are not independent,  i.e., they can be written as functions of  the other variables, specifically $\sigma_{ab}$, $\mathcal{R}_a$ and $\Re_a$. Secondly, the variables ($D_a,Z_a \mathcal{R}_a, \Re_a$) are one index objects, which means that their purely vector part is obtained by taking their $curl$ multiplied by the scale factor. In addition, looking at \rf{ScaVar} above one notices that these variables are in fact gradients of scalars and we know that $\c\3nab_a X=2\omega_a \dot{X} $. This means that at first order one can write:
\begin{eqnarray}
&& (D_a)^V=-2 S^2 \Theta (1+w)\, \omega_a \label{vectD}\,,\\
&& (Z_a)^V=2 S^2 \dot{\Theta}\, \omega_a \label{vectZ}\,,\\
&& (\mathcal{R}_a)^V=2 S^2 \dot{R}\,\omega_a=  4 S^2 \left(\frac{4}{3} \dot{\Theta} \Theta-\frac{2}{3}\frac{K
   }{S^2}\Theta+\ddot{\Theta}\right) \,\omega_a \label{vectR1}\,,\\
&&(\Re_a)^V=2 S^{2} \ddot{R} \,\omega_a= 4 S^2 \left(\frac{8}{9}\frac{ K }{S^2}\Theta^2-\frac{4}{3 }\frac{ K }{S^2}\dot{\Theta }+\frac{8}{3}\dot{ \Theta }^2+\frac{8}{3}\Theta \ddot{\Theta }+2 \Theta ^{(3)}\right) \,\omega_a \label{vectR2}\,,
\end{eqnarray}
i.e., one can express all these quantities in terms of the vorticity vector. Also one can use the constraints(\ref{1+3ShearConstrHO}-
\ref{1+3GrMagDivHO}) to express the remaining quantities $\sigma_{ab}, E_{ab}, H_{ab}$ in terms of the vorticity vector. Thus, in the end the only relevant equations for the dynamics of the vector perturbations is the vorticity equation \rf{1+3VorHO}:
\begin{equation}
 \epsilon  \frac{d\omega_a}{d N}+(2-3w) \omega_a=0\,,
\end{equation}
which does not depend on $\ell$. This means that the vorticity evolution is independent of the scale. Specifically one has
\begin{equation}
\omega^{\ell}= \omega_0^{\ell} \exp\left[-\varepsilon N \; (2-3 w)\right]=\omega_0^{\ell} S^{- \epsilon  (2-3 w)}\,,
 \end{equation}
which implies that the vorticity is always decreasing regardless of the features of the action. This is the same result that one obtains in GR \cite{EBH} and shows that $f(R)$ gravity does not affect the evolution of this quantity. However, the quantities (\ref{vectD}-\ref{vectR2}) will change behavior according to changes in the background and the form of the action.

\subsection{Tensor Perturbations}
As already noticed in \cite{K1}  and shown in Appendix \ref{AppC} the only independent equation in the evolution of the tensor perturbation is the shear. This happens because the  second order equation for $\sigma_{ab}$  is closed and from the constraint  \rf{1+3GrMagConstrHO} one obtains $H_{ab}=(curl \sigma)_{ab}$ where $(\c\,X)^{ab} = \epsilon^{cd\lgl a}\,\3nab_{c}X^{b\rgl}\!_{d}$. In addition to that the first order equation for $\sigma_{ab}$  (Eq. \rf{eqsigma})
can be used to derive $E_{ab}$, because \cite{K1} in $f(R)$ gravity (as in the scalar tensor case \cite{Carloni:2006fs})  the  tensor component  of the anisotropic pressure $\pi_{ab}$ can be proven to be proportional to the shear.
Thus we are left only with the equation for  $\sigma_{ab}$. This equation developed in harmonics and written in terms of the dynamical systems variables reads
 \begin{equation}\label{EqTenIIOrdDynSys}
\frac{d^{2}\sigma^{(\ell)}}{d N^{2}} - \mathcal{A}_2\frac{d \sigma^{(\ell)}}{d N}- \mathcal{B}_2\sigma^{(\ell)}=0 \end{equation}
 \begin{eqnarray}
 &&\mathcal{A}_2=\varepsilon  (2 \chi-2 y+z-\Omega -2)\,,\label{CoeffTens1}\\
&&\mathcal{B}_2=\chi^2-2 (y-z+\Omega -1) \chi+(y-z+\Omega )^2-6 y+5 z-9 \mathcal{K}+(3 w-2) \Omega\,. \label{CoeffTens2}\\
\end{eqnarray}
For \rf{EqTenIIOrdDynSys} the same remark given for scalars holds: since the coefficients depend only on the coordinate of the fixed points two theories with he same fixed point will have, in the fixed point, the same evolution law for the tensor perturbations, like in the case of the scalars. However since the coefficients \rf{CoeffTens1} and  \rf{CoeffTens2} do not contain $\mathds{Q}$ such occurrences is even more common.
\section{Examples}
In this section we will apply the equations defined above to some specific forms of $f(R)$ to illustrate the utility of the above approach.
\subsection{$R^{n}$-gravity}
If  $f(R)=\alpha R^{n}$ we have a model (often called  $R^{n}$-gravity) characterized by the action
\begin{equation}\label{lagrRn}
L=\sqrt{-g}\left[\alpha R^{n}+{\cal L}_{M}\right]\;,
\end{equation}
which constitutes the simplest possible example of fourth order gravity. In $R^n$-gravity  the characteristic function
$\mathds{Q}$ is always constant. In particular, we have
\begin{equation}
\mathds{Q}=\frac{1}{n-1}\,,
\end{equation}
which implies that the variables $z$
and $y$ are not independent, i.e., the phase space of $R^n$-gravity is contained in the subspace $y=n z$ of the general
phase space described by \rf{DynsysNoX}. This can be easily seen if one substitutes $y=n z$ into \rf{DynsysNoX}.
Then the equations for $y$ and $z$ turn out to be exactly the same and
\rf{DynsysNoX} reduces to :
\begin{eqnarray}\label{sistem Rn}
 && \frac{d y}{d N}= n y \varepsilon   \left(\frac{y-2 n (y+2)+5}{(n-1)^2}+\frac{(3-2 n) \chi }{(n-1)^2}-\frac{\Omega }{(n-1)^2}\right)\;,\\
 && \frac{d \Omega}{d N}=-\varepsilon  \Omega  \left(-3 w+\left(3+\frac{2}{n-1}\right) y+3 \chi -\Omega +2\right) \;,\\
 && \frac{d \chi}{d N}=2 \varepsilon   \chi  \left(\chi +1-\frac{n y}{1-n}\right)\;,
 \end{eqnarray}
with the constraint
\begin{equation}\label{constraint materia}
 1+x+y+\chi -\Omega =0\;,
\end{equation}
which is equivalent to the one given in \cite{GenDynSys}. The fixed point with their stability and the associated solutions are given in Tables \ref{FP-R^n}, \ref{SolFP-R^n} and \ref{tabstabRn}. In Table \ref{PertModesRn}, the long wavelength modes of the solutions for the matter scalar perturbations and the tensor perturbations are given. As expected, for the background $t^{2n/3(1+w)}$ corresponding to the point $\mathcal{G}$, the results are the same as the ones already found in \cite{SantePertSca, K1}.
For $\ell\neq 0$, however, one has to use numerical methods to obtain the solution of the equations.

It is interesting to observe the behavior of the matter perturbation modes of the point $\mathcal{F}$. Here the matter perturbations posess a constant mode and the other modes can be growing or decaying depending on the value of the parameter $n$.

As shown in \cite{RnGrav} for some specific values of the parameter $n$ ($1.37\lesssim n\lesssim 2$) this model has a set of cosmic histories characterized by the presence of a transient, decelerated power law expansion that evolves towards an accelerated expansion. Using \rf{EqScaIIOrdDynSys} and \rf{EqTenIIOrdDynSys}  we are now able to see directly the evolution of the scalar and tensor perturbations along these orbits.

\begin{figure}[htbp]
\includegraphics[scale=1.2]{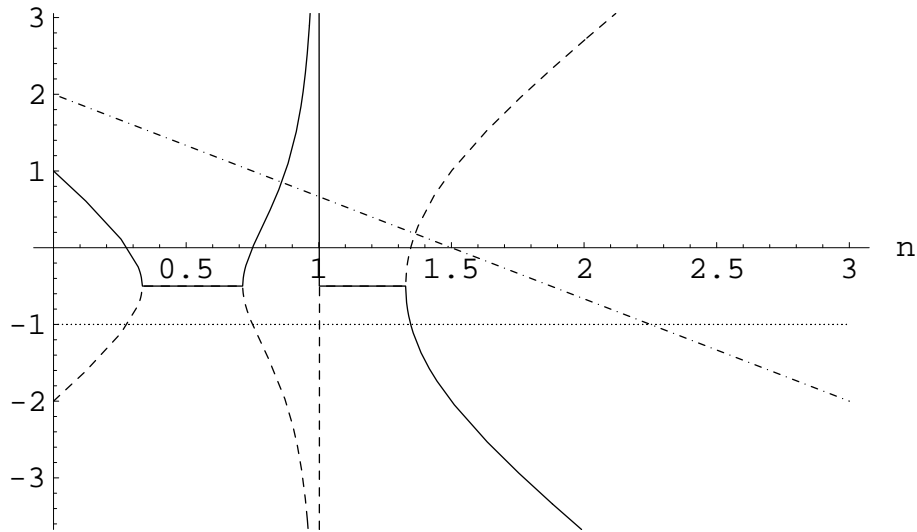}
\caption{Plot against $n$ of the real part of the  exponents of  the long wavelength modes
for $R^n$-gravity in the point $\mathcal{G}$ and in the dust case ($w=0$).}
 \label{Fig1}
\end{figure}

In particular, as the Universe approaches the point $\mathcal{F}$, for $1.37\lesssim n\lesssim 2$  and $w=0$, the large scale scalar perturbations, which nearby  $\mathcal{G}$ have a growing mode (see Figure \ref{Fig1}), start dissipating, which is consistent with what one would expect in a late time acceleration scenario. The large-scale tensor perturbations instead do not change their behavior and keep being dissipated, but at a much faster rate.

In order to analyze the behavior of the perturbations for smaller scales ( $\ell\neq 0$) one needs to  integrate the equations numerically. This can be done in a relative easy manner and an example of the results obtained in the case of dust and  $\ell=100$  are shown in Figures \ref{NumSolF} and \ref{NumSolG}. It is clear that in the point $\mathcal{F}$ the matter scalar perturbations approach a constant value which depends on the initial conditions, while in the point  $\mathcal{G}$  the perturbations first have a phase of growth and then start to decay which is consistent with what was found in \cite{StructForm}. The same can be done with the tensor perturbations, but we find that, as for the $\ell=0$, case these types of perturbations are dissipated on small scales.

\begin{figure}[htbp]
\begin{center}
\includegraphics{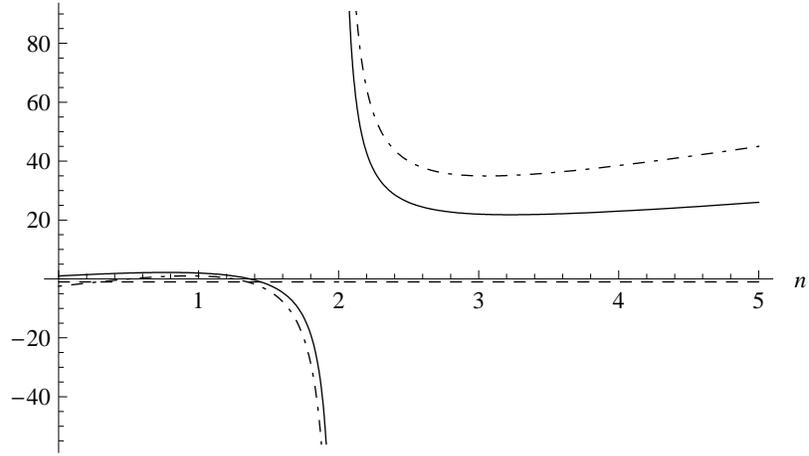}
\caption{Plots against $n$ of the real part of  the exponents of long wavelength matter perturbation modes  for $R^n$-gravity in the point $\mathcal{F}$ in the case of dust ($w=0$).}\label{ModesE}
\label{default}
\end{center}
\end{figure}

\begin{figure}[htbp]
\subfigure[ Normalized plot of   the evolution of $\Delta_m$ in the fixed point $\mathcal{F}$ ]{\includegraphics[scale=0.80
]{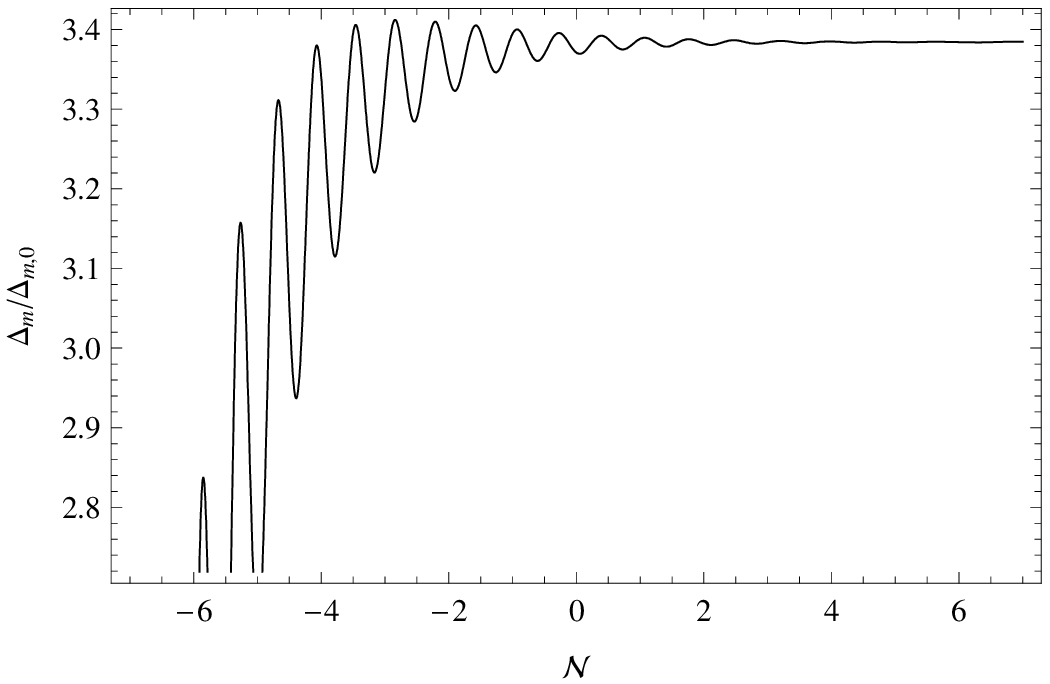}\label{PlotDF}}
\subfigure[ Normalized plot of   the evolution of $\mathds{R}$ in the fixed point $\mathcal{F}$ ]{\includegraphics[scale=0.80]{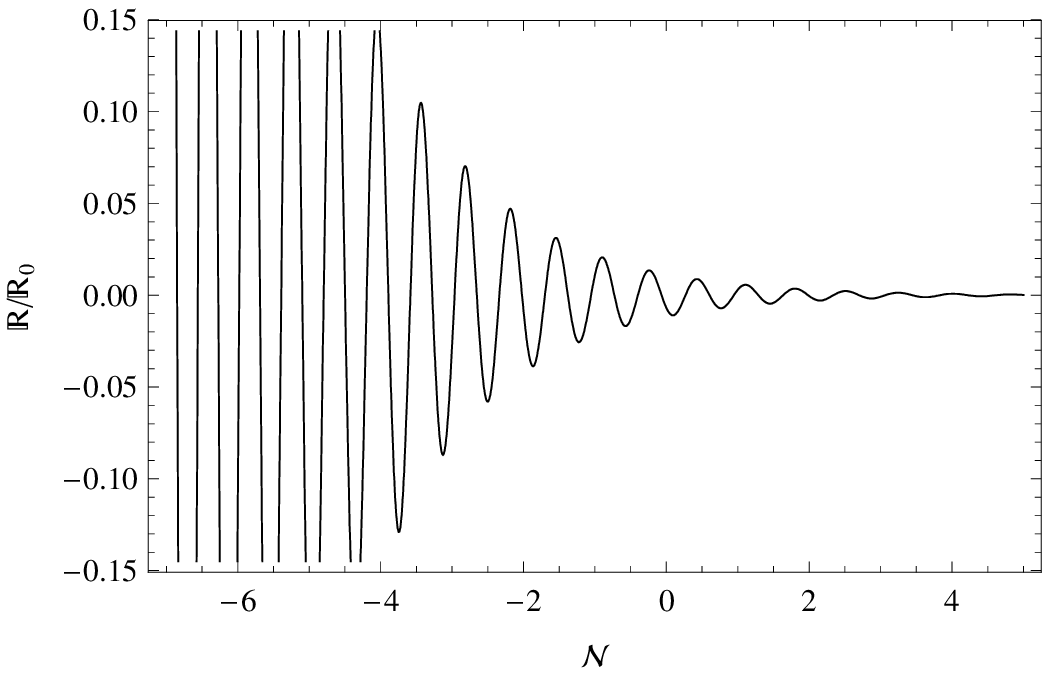}\label{PlotRF}}
\caption{Plots of the solutions of the equations \rf{EqScaIIOrdDynSys}  in the  fixed point $\mathcal{F}$ in the case $n\approx 1.37$, $w=0$ and  $\ell=100$. }\label{NumSolF}
\end{figure}

\begin{figure}[htbp]
\subfigure[ Normalized plot of the evolution of $\Delta_m$ in the fixed point $\mathcal{G}$ ]{\includegraphics[scale=0.80
]{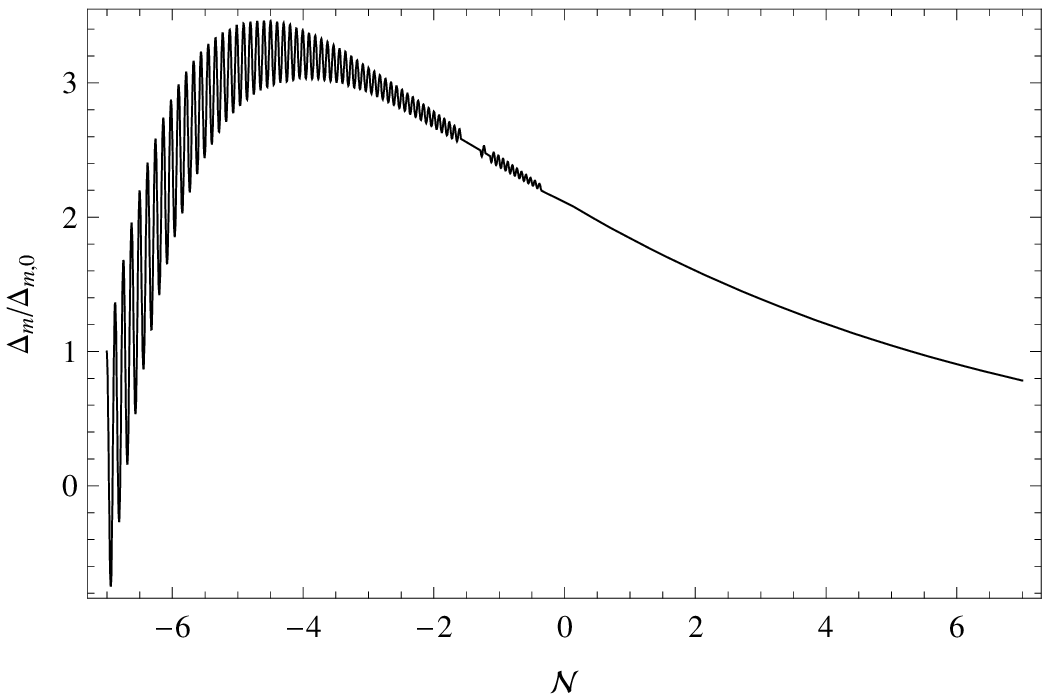}\label{PlotDG}}
\subfigure[ Normalized plot of   the evolution of $\mathds{R}$ in the fixed point $\mathcal{G}$ ]{\includegraphics[scale=0.80]{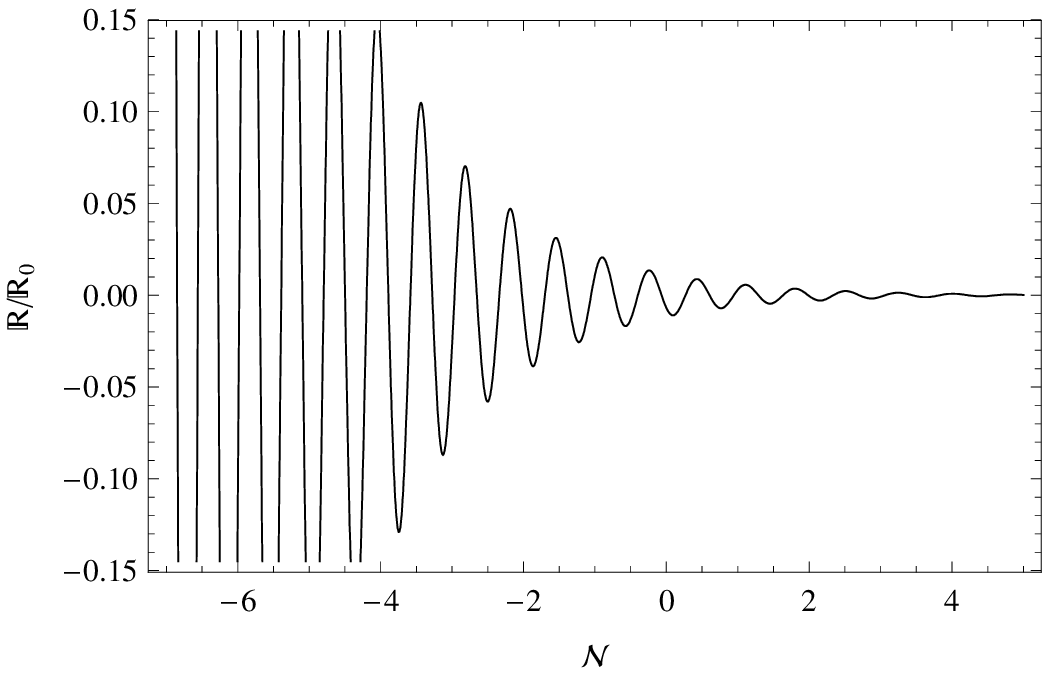}\label{PlotRG}}
\caption{Plots of the solutions of the equations \rf{EqScaIIOrdDynSys}  in the  fixed point $\mathcal{G}$ in the case $n\approx 1.37$, $w=0$ and  $\ell=100$ }\label{NumSolG}
\end{figure}

 \begin{table}[h]
\caption{Coordinates of the fixed points for the model $f(R)\,=\,\alpha R^n$. The
superscript ``*" represents a double solution. The point $\mathcal{B}$ is a
double solution for $n=0,2$.}
\label{FP-R^n}
\begin{tabular}{llrl} \hline\hline
Point & Coordinates $(x,y,z,\Omega)$ & $\chi$ &
\\ \hline
$\mathcal{A}^*$ & $\left(0, 0, 0, 0\right)$ & $-1$ &   \\
$\mathcal{B}$ & $\left(-1, 0, 0, 0\right)$
& $0$ &  \\
$\mathcal{C}$ &
$\left(-1-3w,0,0,-1-3w\right)$ & $-1$ & $$\\
$\mathcal{D}$ & $\left(1-3w,0,0,2-3w\right)$ & $0$ &
$$\\
 $\mathcal{E}$ & $\left(2(1-n),2n(n-1),2(n-1),0\right)$ & $2n(n-1)-1$ & $$\\
$\mathcal{F}$ &  $\left(\frac{2 (n-2)}{2 n-1},\frac{(4 n-5) n}{2
n^2-3 n+1},\frac{4 n-5}{2 n^2-3 n+1},0\right)$ & $0$ & $ $\\
$\mathcal{G}$ & $\left(-\frac{3 (n-1) (w +1)}{n},\frac{4 n-3 (w
-1)}{2 n},\frac{4 n-3 (w-1)}{2 n^2},\frac{-2 (3 w +4)
n^2+(9 w +13) n-3 (w +1)}{2 n^2}\right)$ & $0$& $ $\\ \hline\hline
 \end{tabular}
\end{table}

\begin{table}
\caption{Solutions associated to the fixed points of the model $f(R)\,=\,\alpha R^n$.}
\label{SolFP-R^n}
\begin{tabular}{cccc}
\hline\hline Point & Scale Factor  \\ \hline
$\mathcal{A}$ & $S=S_{0}(t-t_{0})$ & $\mu_m=0$\\
$\mathcal{B}$ & $S=S_{0}(t-t_{0})^{1/2}$ (only for $n=3/2$)&$\mu_m=0$\\
$\mathcal{C}$  & $S=S_{0}(t-t_{0})$ & $\mu_m=0$\\
$\mathcal{D}$ &$S=S_{0}(t-t_{0})^{1/2}$ (only for $n=3/2$) & $\mu_m=0$\\
$\mathcal{E}$ & $\left\{ \begin{array}{ccc} S=\frac{K t}{2n^{2}-2n-1}
& \mbox{if}&  K \neq 0\\ S=S_{0}t & \mbox{if}& K=0 \end{array}\right.$&$\mu_m=0$\\
$\mathcal{F}$  & $S=S_{0}\;t^{\frac{(1-n)(2n-1)}{n-2}}$ &$\mu_m=0$\\
$\mathcal{G}$ &$S=S_{0}\;
t^{\frac{2n}{3(1+w)}}$& $\mu_m=\mu_{m,0} t^{-2n}$ \\  &
$\mu_{m, 0}=(-1)^{n}3^{-n}2^{2n-1} n^{n} (1+w)^{-2n}\times$\\
&$(4n-3(1+w))^{n-1} [2n^{2}(4+3w)-n(13+9w)+3(1+w)]$\\  \hline\hline \\
\end{tabular}
\label{tavola punti fissi materia 1}
\end{table}

\begin{table}
\caption{Stability of the fixed points for $R^{n}$-gravity with
matter. We consider here only dust, radiation. The term
``spiral$^{+}$" has been used for pure attractive focus-nodes and  the
term``saddle-focus" for unstable focus-nodes. }\label{tabstabRn} \centering
\bigskip
\begin{tabular}{ccccc}
\hline\hline & $n<\frac{1}{2}(1-\sqrt{3})$ & $\frac{1}{2}(1-\sqrt{3})<n<0$
& $0<n<1/2$& $1/2<n<1$
     \\ \hline
 $\mathcal{A}$ & saddle& saddle& saddle& saddle\\
 $\mathcal{B}$  & repellor& repellor& repellor& repellor \\
 $\mathcal{C}$ & saddle& saddle& saddle& saddle\\
 $\mathcal{D}$ & saddle& saddle& saddle& saddle\\
 $\mathcal{E}$ & saddle& attractor &  spiral &  spiral\\
 $\mathcal{F}$ & attractor& saddle& saddle& attractor \\
\hline\hline \\
\end{tabular}
\begin{tabular}{ccccc}
  \hline\hline  &$1<n<5/4$& $5/4<n<4/3$ & $4/3<n<\frac{1}{2}(1+\sqrt{3})$ & $n>\frac{1}{2}(1+\sqrt{3})$ \\ \hline
 $\mathcal{A}$ & saddle& saddle& saddle& saddle\\
 $\mathcal{B}$   & saddle& repellor& repellor& repellor\\
 $\mathcal{C}$ & saddle& saddle& saddle& saddle\\
 $\mathcal{D}$ &saddle& saddle& saddle& saddle\\ 
 $\mathcal{E}$  &spiral&  spiral & attractor & saddle \\
 $\mathcal{F}$ &repellor&saddle&saddle& attractor\\
\hline\hline\\
\end{tabular}
\centering
\begin{tabular}{ccccccc}
\hline\hline
 $\mathcal{G}$& $n\lesssim 0.33$ & $0.33\lesssim n\lesssim 0.35$
& $0.35\lesssim n\lesssim 0.37$& $0.37\lesssim n\lesssim 0.71$&
  $0.71\lesssim n\lesssim 1$ \\\hline
 $w=0$  &saddle&saddle-focus&saddle-focus&saddle-focus&saddle \\
 $w=1/3$ &saddle&saddle&saddle-focus&saddle-focus&saddle-focus\\
\end{tabular}
\begin{tabular}{ccccccc}
\hline &   $1\lesssim n\lesssim 1.220$ & $1.220\lesssim n\lesssim
1.223$& $1.223\lesssim n\lesssim 1.224$& $1.224\lesssim n\lesssim
1.28$ \\\hline
 $w=0$  &saddle-focus&saddle-focus&saddle-focus &saddle-focus\\
 $w=1/3$ &saddle-focus&saddle-focus&saddle-focus&saddle-focus\\
\end{tabular}
\begin{tabular}{ccccccc}
\hline& $1.28\lesssim n\lesssim 1.32$& $1.32\lesssim n\lesssim 1.47$&
$1.47\lesssim n\lesssim 1.50$& $n\gtrsim 1.50$  \\\hline
 $w=0$ &saddle-focus&saddle& saddle & saddle \\
 $w=1/3$&saddle&saddle& saddle & saddle \\\hline\hline
\end{tabular}
\end{table}


\begin{table}[h]
\caption{Exponent of the modes for scalar and tensor in the fixed points of the system \rf{sistem Rn}. }
\label{PertModesRn}
\begin{tabular}{lll} \hline\hline
Point & Scalar Modes Exponents& Tensor Modes Exponents
\\ \hline
$\mathcal{A}$ &$\{- 2\varepsilon ,0\}$ & $\left\{ -\varepsilon (2+\sqrt{3}) , -\varepsilon (2-\sqrt{3})\right\}$   \\
$\mathcal{B}$ &$\left\{0,\varepsilon ,\varepsilon \left(\frac{3w}{2}-1\right)\right\}$ & $\left\{ -\varepsilon , 0 \right\}$  \\
$\mathcal{C}$ &
$\{-2\varepsilon ,0,\varepsilon (3 w-1),\varepsilon (3 w+1)\}$ & $\left\{\frac{\varepsilon}{2} \left(3 w-\sqrt{9 w^2-6 w+9}-3\right),\right.$\\&&$\left.\frac{\varepsilon}{2} \left(3 w+\sqrt{9 w^2-6 w+9}-3\right)\right\}$   \\
$\mathcal{D}$ & $\left\{0,\varepsilon , \varepsilon \left(\frac{3 w}{2}-1)\right)\right\}$ &$\left\{ 0 , \varepsilon \left(\frac{3 w}{2}-2\right)\right\}$     \\
 $\mathcal{E}$ & $\left\{0,\varepsilon (3 w-1), \varepsilon \left[n-2-\sqrt{3} \sqrt{n (3 n-4)}\right],\right.$\\
&$\left. , \varepsilon \left[n-2+\sqrt{3} \sqrt{n (3 n-4)}\right]\right\}$&$\left\{\varepsilon \left(n-3-\sqrt{6-3 n^2}\right) , \varepsilon \left(n-3+\sqrt{6-3 n^2}\right) \right\}$  \\
$\mathcal{F}$ & $\left\{0,\varepsilon \left(8 n+2+\frac{9}{n-2}\right),\varepsilon \left[4 (n+1)+\frac{6}{n-2}\right]\right.$,
\\&$\left.-\varepsilon \left[ \frac{3 (n (2 n-3)+1) w}{n-2}+1\right]\right\}$ & $\left\{\varepsilon n \left(4+\frac{3}{n-2}\right) , 8 \varepsilon \left(n+1+\frac{9}{n-2}\right)\right\}$ \\
$\mathcal{G}$ & $\left\{\varepsilon \left(2-\frac{4 n}{3 (w+1)}\right),\varepsilon \left(\frac{2 n w}{w+1}-1\right),\right.$\\&$\left.\varepsilon \left[\frac{3 (w+1)+3 n ((2 n-3) w-1)+ \sqrt{A}}{6 (n-1) (w+1)}\right],\right.$\\&$\left.\varepsilon \left[\frac{3 (w+1)+3 n ((2 n-3) w-1)-\sqrt{A}}{6 (n-1) (w+1)}\right]\right\}$ & $\left\{ \varepsilon \left(1-\frac{4 n}{3 (w+1)}\right), \varepsilon \frac{2 (n-1) w-2}{w+1}\right\}$    \\ \hline\multicolumn{3}{c}{$A=(n-1)(4 (3 w+8)^2 n^3-4 (3
   w (18 w+55)+152) n^2+3 (w+1) (87 w+139) n-81 (w+1)^2)$}  \\ \hline\hline
 \end{tabular}
\end{table}
\section{The case $f(R)=R+\alpha R^{n}$}
In this case the action reads
\begin{equation}\label{lagr RRn}
L=\sqrt{-g}\left[R+\alpha R^{n}+{\cal L}_{M}\right]\;.
\end{equation}
This theory has gained much popularity as a fourth order gravity model within the
context of both inflation and dark energy \cite{Ottewill, star80, Suen, Carroll}. The characteristic function
$\mathds{Q}$ is \,:
\begin{equation}\label{m-R+Rngen}
\mathds{Q}\,=\,\frac{y}{n(z-y)}\,,
\end{equation}
and substituting this relation into the system of equations \rf{DynsysNoX} we obtain
\begin{eqnarray}\label{sistem R+Rn}
 && \frac{d y}{d N}=\varepsilon  y \left[2 \chi-2 y+\frac{y(-\chi+y-z+\Omega -1)}{n y-n z}+4\right]\;,\\
 && \frac{d z}{d N}=\varepsilon\left[\frac{ (-\chi+y-z+\Omega-1) y^2}{n y-n z}+  z (3\chi-3 y+z-\Omega+5)\right]\;,\\
 && \frac{d \Omega}{d N}=-\varepsilon  \Omega(3 w-3 \chi+3 y-z+\Omega-2)\;,\\
 && \frac{d \chi}{d N}=2 \varepsilon  \chi(\chi-y+1)\;,
 \end{eqnarray}
with the constraint
\begin{equation}
 1+x-y+z+\chi=0\;,
\end{equation}
The  fixed points and their stability for the phase space of  \rf{sistem R+Rn} are  shown in Tables \ref{fixpointR+Rn} \ref{tabRRnSol} and \ref{tabR+RnStab}.  In Table \ref{PertModesRRn} we show the exponent of the long wavelength scalar and tensor perturbation modes.

The first point to note is that there exists a fixed point that corresponds to a transient Friedmann type behavior just as in the $f(R)=R^n$ case for exactly the same values of the dynamical system variables. It is therefore not surprising that the long wavelength scalar perturbations modes for these fixed points have the same solutions in both theories. At first glance this may seem to contradict the results in \cite{StructForm}, where it was found that the scalar perturbations at the fixed point $\mathcal{L}$ depend on the value of $\alpha$. However, this discrepancy can be resolved when one considers the structure of our dynamical system formalism and in particular the conditions \rf{FPcond}. These additional equations impose further constraints on our system. This can be seen more clearly if we consider the fixed point $\mathcal{L}$ in the case of dust ($w=0$) in more detail. Using the definition of the dynamical systems variable, $x$ and our choice of $f(R)$ we find that the following constraint must be obeyed at $\mathcal{L}$
\begin{equation}\label{condX}
 x=-\frac{3 (n-1)}{n}\qquad\Rightarrow\qquad \frac{S}{3\dot{S}}\frac{\alpha  n^2 \dot{R} R^{(n-1)}}{R + \alpha n R^n}=-1\;.
\end{equation}
If we then use the fact that
\begin{eqnarray}
R=6\left(\frac{\ddot{S}}{S}+\frac{\dot{S}^2}{S^2}+\frac{K}{S^2}\right), \qquad S(t)= S_{0} \left(t-t_{0}\right)^{\frac{2n}{3}}
\end{eqnarray}
and that we require that the constraint above is satisfied for all time, it is trivial to show that  \rf{condX} is only satisfied if one requires $\alpha \rightarrow \infty$. This means that supposing that the point $\mathcal{L}$ is associated with a solution of the cosmological equations, means that the function $f(R)$ we are dealing with is very close to $R^n$. As a consequence, when we insert constant values of the dynamical system variables in the coefficient of the perturbation equations we are in fact  imposing that we are dealing with a theory which is essentially $R^n$-gravity. Thus, naturally, we recover exactly the same results of the previous section. This can be also seen from the fact that if we substitute the coordinates of the fixed point in the definition of $\mathds{Q}$ we obtain
\begin{equation}
\mathds{Q}=\frac{1}{1-n}\,,
\end{equation}
which is exactly the $\mathds{Q}$ of $R^n$-gravity. Since $\mathds{Q}$ is the only parameter that differentiate the theory of gravity in the coefficients of the perturbation equations (as well as the dynamical system) we clearly expect that after substituting the coordinates of the fixed point into the coefficients we obtain the same results of $R^n$-gravity.

This seems to suggest that in some sense the fixed points carry information about the theory of gravity other than the background evolution, so that a fixed point represents a physical solution only of a specific form of the Lagrangian. Hence if the phase space of a generic theory of gravity posses that fixed point it means that there can exist regimes in which this theory can be approximated by a Lagrangian for which that background is a ``physical" solution.

Furthermore, from what was said above, one can also conclude that somehow the evolution of the perturbations is attached to the fixed point in such a way that regardless of the theory, the evolution of scalar perturbations is determined only by the fixed point. In other words we are somehow obtaining the ``fixed point" of the perturbation theory which corresponds to the fixed point of the dynamical systems approach. Such a fixed point is an approximation of the real behavior of the equations. This means that we expect the scalar perturbations around this fixed point to be approximated by the results we found, which is consistent with what we obtained in \cite{StructForm}.

\begin{table}[h]
\caption{Coordinate of the finite fixed points for $R+\alpha
R^{n}$ gravity.\label{fixpointR+Rn}}
\begin{tabular}{llrl} \hline\hline
Point & Coordinates $(x,y,z,\Omega)$ & $\chi$ &
\\ \hline
$\mathcal{A}$ & $\left(0, 0, 0, 0\right)$ & $-1$ &   \\
$\mathcal{B}$ & $\left(-1, 0, 0, 0\right)$
& $0$ &  \\
$\mathcal{C}$ &
$\left(-1-3w,0,0,-1-3w\right)$ & $-1$ & $$\\
$\mathcal{D}$ & $\left(1-3w,0,0,2-3w\right)$ & $0$ &
$$\\
$\mathcal{E}$ & $\left(0, 2, 1, 0\right)$& $0$&
 \\
$\mathcal{F}$ & $\left(2,0,-2,0\right)$ & $-1$ & $$
 \\
 $\mathcal{G}$ & $\left(4,0,-5,0\right)$ & $0$ & $$ \\
 $\mathcal{H}$ & $\left(2(1-n),2n(n-1),2(1-n),0\right)$ & $2n(n-1)-1$ & $$\\
$\mathcal{I}$ & $\left(\frac{2 (n-2)}{2 n-1},\frac{(5-4 n) n}{2
n^2-3 n+1},\frac{5-4 n}{2 n^2-3 n+1},0\right)$ & $0$ & $ $ \\
$\mathcal{L}$& $\left(-\frac{3 (n-1) (w +1)}{n},\frac{-4 n+3 w +3}{2
n},\frac{-4 n+3w +3}{2 n^2},\frac{-2 (3 w +4) n^2+(9 w +13) n-3 (w
+1)}{2 n^2}\right)$ & $0$& $ $ \\ \hline\hline
 \end{tabular}
\end{table}

\begin{table}[ht]
\caption{Solutions associated to the fixed points of $R+\alpha R^n$.
The solutions are physical only in the intervals of $p$ mentioned in
the last column.} \label{tabRRnSol}
\begin{tabular}{lccc} \hline\hline
Point & Scale Factor & Energy Density& Physical
\\ \hline
$\mathcal{A}$ & $S=  S_0\left(t-t_0\right)$ & $0$ &  $n\geq 1$\\
$\mathcal{B}$ & $S= S_0 \left(t-t_0\right)^{1/2}$ & $0$&  $n\geq1$\\
$\mathcal{C}$ &$S= S_0\left(t-t_0\right)$& $0$&$n\geq1$ \\
$\mathcal{D}$ &$S= S_0 \left(t-t_0\right)^{1/2}$& $0$&  $n\geq1$\\
$\mathcal{E}^{*}$ & $\begin{array}{c}\left\{
                      \begin{array}{l}
                        S= S_0,  \\
                        S=S_0\exp\left[\pm
                        2\sqrt{3}\alpha^{\gamma}(2-3n)^{\gamma}(t-t_{0})\right],\quad
                        \end{array}\right.\\\gamma=\frac{1}{2(1-n)} \end{array}$&0&$\begin{array}{c}
                                     n\geq0\\
                                    \begin{array}{c}
                                      n<\frac{2}{3}, \alpha>0 \;\; \vee\\
                                      n>\frac{2}{3}, \alpha<0
                                    \end{array}
                                   \end{array}$\\
$\mathcal{F}$ & $S= \left(t-t_0\right)$& $0$&$n\geq1$ \\
$\mathcal{G}$ & $S= S_0 \left(t-t_0\right)^{1/2}$ & 0 &  $n\geq1$\\
$\mathcal{H}$ & $S=  \sqrt{1-2n(n-1)}\left(t-t_0\right)$ & $0$&$1\le n\geq \frac{1}{2}+\frac{\sqrt{3}}{2}$\\
$\mathcal{I}^{*}$ & $S=S_0\left(t-t_0\right)^{\frac{2 n^2-3
n+1}{2-n}}$  & $\mu^m=\mu_{m\,0}t^{-\frac{3 \left(2 n^2-3 n+1\right) (w +1)}{n-2}}$& $n=\frac{1}{2}, \mu_{m\,,0}=0$\\
$\mathcal{\mathcal{L}}$& $S= S_0\left(t-t_0\right)^{\frac{2 n}{3
(w
+1)}}$  & $\mu_m=\mu_{m\,,0} (t-t_{0})^{2 p} $ & non physical\\
\hline\hline
 \end{tabular}
\end{table}

\begin{table}[ht] \centering \caption{The stability of  the fixed points in the model $R+\alpha R^n$. The quantities
$B_i$ related to the fixed point $\mathcal{L}$, represent some non
fractional numerical values ($B_1\approx1.220$, $B_1\approx1.224$,
$B_3\approx1.470$). \label{tabR+RnStab}}
\begin{tabular}{ll}
\hline\hline Point &  Stability \\ \hline
& \\
$\mathcal{B}$ &
saddle  \\
$\mathcal{B}$ & $\left\{\begin{array}{cc}
                  \mbox{repellor} &  0<w<2/3\\
                  \mbox{saddle} & \mbox{otherwise}
                \end{array}\right.$\\
$\mathcal{C}$ & saddle  \\
$\mathcal{D}$ & $\left\{\begin{array}{cc}
                  \mbox{repellor} &  2/3<w<1\\
                  \mbox{saddle} & \mbox{otherwise}
                \end{array}\right.$\\
$\mathcal{E}$ & $\left\{\begin{array}{cc}
                  \mbox{attractor} &  \frac{32}{25}\le n < 2\\
                  \mbox{spiral}^+ &  0 < n < \frac{32}{25}\\
                  \mbox{saddle} & \mbox{otherwise}
                \end{array}\right.$\\
$\mathcal{F}$ & saddle  \\
$\mathcal{G}$ & saddle \\
$\mathcal{H}$ & $\left\{\begin{array}{cc}
                  \mbox{attractor} & \frac{1}{2}(1-\sqrt{3})<n\le 0\\
                  \mbox{spiral}^+ &  0<n<1\\
                  \mbox{saddle} & \mbox{otherwise}
                \end{array}\right.$\\
$\mathcal{I}$ & $\left\{\begin{array}{lc}
 \mbox{attractor}   & n<\frac{1}{2}(1-\sqrt{3})\cup n>2 ,  \\
                                      \mbox{repeller}  & \left\{\begin{array}{cc}
                                                                  & 1<n<\frac{5}{4},  (w=0,1/3),\\
                                                                  &1<n<\frac{1}{14}(11+\sqrt{37}),  (w=1)
                                                                   \end{array}\right. \\
                                      \mbox{saddle}      &  \mbox{otherwise},\\
                \end{array}\right.$\\
$\mathcal{L}$ &  $\left\{\begin{array}{lc}
                  w=0,1/3   & \mbox{saddle}, \\
                  w=1   & \left\{\begin{array}{cc}
                  \mbox{repellor} &  B_1<n\le B_2\cup B_3<n<\frac{3}{2} ,\\
                  \mbox{saddle} & \mbox{otherwise}
                \end{array}\right.
                \end{array}\right.$\\\hline\hline
\end{tabular}
\end{table}

\begin{table}[h]
\caption{Exponent of the modes for scalar and tensor in the fixed points of the system \rf{sistem R+Rn}. }
\label{PertModesRRn}
\begin{tabular}{lll} \hline\hline
Point & Scalar Modes Exponents& Tensor Modes Exponents
\\ \hline
$\mathcal{A}$ &$\{- 2\varepsilon ,0\}$ & $\left\{ -\varepsilon (2+\sqrt{3}) , -\varepsilon (2-\sqrt{3})\right\}$   \\
$\mathcal{B}$ &$\left\{0,\varepsilon ,\varepsilon \left(\frac{3w}{2}-1\right)\right\}$ & $\left\{ -\varepsilon , 0 \right\}$  \\
$\mathcal{C}$ &
$\{-2\varepsilon ,0,\varepsilon (3 w-1),\varepsilon (3 w+1)\}$ & $\left\{\frac{\varepsilon}{2} \left(3 w-\sqrt{9 w^2-6 w+9}-3\right),\right.$\\&&$\left.\frac{\varepsilon}{2} \left(3 w+\sqrt{9 w^2-6 w+9}-3\right)\right\}$   \\
$\mathcal{D}$ & $\left\{0,\varepsilon , \varepsilon \left(\frac{3 w}{2}-1)\right)\right\}$ &$\left\{ 0 , \varepsilon \left(\frac{3 w}{2}-2\right)\right\}$     \\
$\mathcal{E}$ &$\{\varepsilon\frac{-3 n-\sqrt{25 n-32} \sqrt{n}}{2 n},\varepsilon\frac{-3 n+\sqrt{25 n-32} \sqrt{n}}{2 n}\}$ & $\left\{ 0 , 0\right\}$   \\
$\mathcal{F}$ &$\left\{-2\varepsilon,-2\varepsilon, 0, -\varepsilon(1-3w)\right\}$ & $\left\{ \varepsilon(-3-\sqrt{6}) , \varepsilon(-3+\sqrt{6}) \right\}$  \\
$\mathcal{G}$ &
$\{-5\varepsilon ,0, 2\varepsilon,\varepsilon (3 w-2)\}$ & $\left\{- \frac{7}{2} \varepsilon,0 \right\}$   \\
 $\mathcal{H}$ & $\left\{0,\varepsilon (3 w-1), \varepsilon \left[n-2-\sqrt{3} \sqrt{n (3 n-4)}\right],\right.$\\
&$\left. , \varepsilon \left[n-2+\sqrt{3} \sqrt{n (3 n-4)}\right]\right\}$&$\left\{\varepsilon \left(n-3-\sqrt{6-3 n^2}\right) , \varepsilon \left(n-3+\sqrt{6-3 n^2}\right) \right\}$  \\
$\mathcal{I}$ & $\left\{0,\varepsilon \left(8 n+2+\frac{9}{n-2}\right),\varepsilon \left[4 (n+1)+\frac{6}{n-2}\right]\right.$,
\\&$\left.-\varepsilon \left[ \frac{3 (n (2 n-3)+1) w}{n-2}+1\right]\right\}$ & $\left\{\varepsilon n \left(4+\frac{3}{n-2}\right) , 8 \varepsilon \left(n+1+\frac{9}{n-2}\right)\right\}$ \\
$\mathcal{L}$ & $\left\{\varepsilon \left(2-\frac{4 n}{3 (w+1)}\right),\varepsilon \left(\frac{2 n w}{w+1}-1\right),\right.$\\&$\left.\varepsilon \left[\frac{3 (w+1)+3 n ((2 n-3) w-1)+ \sqrt{A}}{6 (n-1) (w+1)}\right],\right.$\\&$\left.\varepsilon \left[\frac{3 (w+1)+3 n ((2 n-3) w-1)-\sqrt{A}}{6 (n-1) (w+1)}\right]\right\}$ & $\left\{ \varepsilon \left(1-\frac{4 n}{3 (w+1)}\right), \varepsilon \frac{2 (n-1) w-2}{w+1}\right\}$    \\ \hline\multicolumn{3}{c}{$A=(n-1)(4 (3 w+8)^2 n^3-4 (3
   w (18 w+55)+152) n^2+3 (w+1) (87 w+139) n-81 (w+1)^2)$}  \\ \hline\hline
 \end{tabular}
\end{table}

\section{Conclusions}

In this paper we have discussed the connection between the dynamical system approach and the covariant-gauge invariant theory of perturbations,
presenting a method to calculate directly the evolution of the scalar, vector and tensor perturbations at a fixed point of the phase space of a generic $f(R)$-gravity theory.
Within the limitations of the dynamical system approach one is then able to obtain an idea of the evolution of the perturbations in any $f(R)$ model.

Because of the non-linearity and the peculiar structure of the dynamical system formalism, the concept of fixed points in $f(R)$-gravity is more subtle than the one of GR. In particular, one can have fixed points which do not correspond to solutions which satisfy the cosmological equations. This is due to the fact that the fixed point conditions \rf{FPcond} constitute additional constraints that can be incompatible with the cosmological equations and that the exact solution associated to the fixed points seems to depend more on the definition of the variables than the model itself.

This has profound implications on the interpretation of the results of the dynamical system approach. Specifically it suggests that this kind of approach offers only partial information on the actual evolution of the cosmology in this framework and that this information depends on the formalism used. In particular (i) one might not be able to see all the fixed points in the phase space because of the form of the variables  and (ii) some of these fixed points might not correspond to actual solutions of the system with obvious problems in the interpretations of the orbits which have these points as attractors. As a consequence one has to be very careful in using these tools to derive general conclusions about the dynamics for these cosmological models.

These peculiarities also have consequences on the results of the perturbations equations. In particular, we found that the evolution laws obtained using this form of the equations do not in general coincide with those that one would obtain by simply subsittuting in the background corresponding to the fixed point in the dynamical system equations. This can be explained when one considers that the additional conditions associated with the fixed points, further constrain the perturbation equations and therefore lead to different results. On the other hand such a fact also implies that one can associate an evolution law for first order perturbations to a specific fixed point,  which  in some sense may be  considered  as  a ``fixed point" for the perturbation theory. Thus one can use the results of the equations given above in the same way in which one used the solution at the fixed points: gaining qualitative information about the behavior of the perturbations along an orbit. Such a feature, confirmed by the direct calculations presented in \cite{StructForm}, can help with the understanding of the behavior of the perturbations in models for which a direct numerical integration is to complex or to resource intensive to be performed. It is also worth noticing that since the peculiarities in the predictions of the equations presented above derive ultimately from the additional condition on the fixed points, they only apply when one deals with fixed points and will not be present when one considers the dynamical system variables as functions of time.

In conclusion, in spite of these difficulties the dynamical system approach, when combined with the covariant gauge invariant formalism, represents an extremely powerful tool for the study of $f(R)$-gravity cosmological models. We believe that a careful use of these methods could be invaluable in determining the relevance of these models on cosmological scales and to constrain them using current and future observations.

\acknowledgments
SC was founded by the Generalitat de Catalunya. PKSD, KA and MA are funded by the NRF (South Africa).

\appendix
\section{General propagation and constraint equations of the 1+3 covariant formalism.}\label{CovID}

In this appendix we will write the propagation and constraint equations for the kinematical  and thermodynamical quantities of the 1+3 formalism for a generic $f(R)$-theory of gravity. Chosen the frame $u_a$ the kinematical quantities are defined as
\begin{equation}
\Theta=\3nab^a u_a\,, \qquad  \sigma_{ab}=\3nab_{(a} u_{b)}\,,\qquad  \omega_{ab}=\3nab_{[a} u_{b]}\,,\qquad
\dot{u}_{a}=u_c\nabla^{c} u_{a}\,,
\end{equation}
and the thermodynamical ones as in \rf{mupitot} and \rf{qpaitot} with the condition
\begin{equation}
q_{a}=q_{\langle{a}\rangle}\,,\qquad\pi_{a b }=\pi_{\langle{a}{b}\rangle}\,.
\end{equation}
As usual angle brackets applied to a vector  denote
the projection of this vector on the tangent 3-spaces
\begin{equation}
V_{\langle{{a}}\rangle}=h_{{a}}{}^{{b}} V_{{b}}\;.
\end{equation}
Instead when applied to a tensor they denote the projected,
symmetric and trace free part of this object
\begin{equation}
W_{\langle{{a}}{{b}}\rangle}=\left[h_{({{a}}}{}^{c}
h_{{{b}})}{}^{d}-
{\textstyle\frac{1}{3}}h^{{c}{d}}h_{{{a}}{{b}}}\right]W_{{c}{d}}\,.
\end{equation}

\noindent Expansion propagation (generalized Raychaudhuri equation):
\begin{eqnarray}\label{1+3eqRayHO}
&&\dot{\Theta}+{\textstyle\frac{1}{3}}\Theta^2+\sigma_{{{a}}{{b}}}
\sigma^{{{a}}{{b}}} -2\omega_{{a}}\omega^{{a}} -\tilde{\nabla}^a
\dot{u}_{{a}}+ \dot{u}_{{a}}
\dot{u}^{{a}}+{\textstyle\frac{1}{2}}(\tilde{\mu}^{m} +
3\tilde{p}^{m}) =-{\textstyle\frac{1}{2}}({\mu}^{R} +
3{p}^{R})\;.
\end{eqnarray}
Vorticity propagation:
\begin{equation}\label{1+3VorHO}
\dot{\omega}_{\langle {{a}}\rangle }
+{\textstyle\frac{2}{3}}\Theta\omega_{{a}} +{\textstyle\frac{1}{2}}\curl
\dot{u}_{{a}} -\sigma_{{{a}}{{b}}}\omega^{{b}}=0 \;.
\end{equation}
Shear propagation:
\begin{equation}\label{1+3ShearHO}
\dot{\sigma}_{\langle {{a}}{{b}} \rangle }
+{\textstyle\frac{2}{3}}\Theta\sigma_{{{a}}{{b}}}
+E_{{{a}}{{b}}}-\D_{\langle {{a}}}\dot{u}_{{{b}}\rangle }
+\sigma_{{c}\langle {{a}}}\sigma_{{{b}}\rangle }{}^{c}+
\omega_{\langle {{a}}}\omega_{{{b}}\rangle} - \dot{u}_{\langle
{{a}}}\dot{u}_{{{b}}\rangle}
\,=\,{\textstyle\frac{1}{2}}\pi^{R}_{{{a}}{{b}}}\;.
\end{equation}
Gravito-electric propagation:
\begin{eqnarray}\label{1+3GrElHO}
 && \dot{E}_{\langle {{a}}{{b}} \rangle }
+\Theta E_{{{a}}{{b}}} -\curl H_{{{a}}{{b}}}
+{\textstyle\frac{1}{2}}(\tilde{\mu}^{m}+\tilde{p}^{m})\sigma_{{{a}}{{b}}}
-2\dot{u}^{c}\ep_{{c}{d}({{a}}}H_{{{b}})}{}^{d} -3\sigma_{{c}\langle
{{a}}}E_{{{b}}\rangle }{}^{c} +\omega^{c}
\ep_{{c}{d}({{a}}}E_{{{b}})}{}^{d}
\nonumber\\&&~~{}=-{\textstyle\frac{1}{2}}(\mu^{R}+p^{R})\sigma_{{{a}}{{b}}}
-{\textstyle\frac{1}{2}}\dot{\pi}^{R}_{\langle {{a}}{{b}}\rangle  }
-{\textstyle\frac{1}{2}}\D_{\langle {{a}}}q^{R}_{{{b}}\rangle }
-{\textstyle\frac{1}{6}}
\Theta\pi^{R}_{{{a}}{{b}}}-{\textstyle\frac{1}{2}}\sigma^{c}{}_{\langle
{{a}}}\pi^{R}_{{{b}}\rangle {c}} -{\textstyle\frac{1}{2}}
\omega^{c}\ep_{{c}({{a}}}^{d}\pi^{R}_{b )d}\;.
\end{eqnarray}
Gravito-magnetic propagation:
\begin{eqnarray}\label{1+3GrMagHO}
 &&\dot{H}_{\langle
{{a}}{{b}} \rangle } +\Theta H_{{{a}}{{b}}} +\curl E_{{{a}}{{b}}}-
3\sigma_{{c}\langle {{a}}}H_{{{b}}\rangle }{}^{c} +\omega^{c}
\ep_{{c}{d}({{a}}}H_{{{b}})}{}^{d}
+2\dot{u}^{c}\ep_{{c}{d}({{a}}}E_{{{b}})}{}^{d}
\nonumber\\&&~~{}={\textstyle\frac{1}{2}}\curl\pi^{R}_{{{a}}{{b}}}-{\textstyle\frac{3}{2}}\omega_{\langle
{{a}}}q^{R}_{{{b}}\rangle
}+{\textstyle\frac{1}{2}}\sigma^{c}{}_{({{a}}}
\ep_{{{b}}){c}}^{\;\;\;\;d}q^{R}_{d}\;.
\end{eqnarray}
Vorticity constraint:
\begin{equation}\label{1+3VorConstrHO}
\D^{{a}}\omega_{{a}} -\dot{u}^{{a}}\omega_{{a}} =0\;.
\end{equation}
Shear constraint:
\begin{equation}\label{1+3ShearConstrHO}
\D^{{b}}\sigma_{{{a}}{{b}}}-\curl\omega_{{a}}
-{\textstyle\frac{2}{3}}\D_{{a}}\Theta +2[\omega,\dot{u}]_{{a}} =
-q^{R}_{a}\;.
\end{equation}
Gravito-magnetic constraint:
\begin{equation}\label{1+3GrMagConstrHO}
 \curl\sigma_{{{a}}{{b}}}+\D_{\langle {{a}}}\omega_{{{b}}\rangle  }
 -H_{{{a}}{{b}}}+2\dot{u}_{\langle {{a}}}
\omega_{{{b}}\rangle  }=0 \;.
\end{equation}
Gravito-electric divergence:
\begin{eqnarray}\label{1+3GrElConstrHO}
&& \D^{{b}} E_{{{a}}{{b}}}
-{\textstyle\frac{1}{3}}\D_{{a}}\tilde{\mu}^{m} -[\sigma,H]_{{a}}
+3H_{{{a}}{{b}}}\omega^{{b}}={\textstyle\frac{1}{2}}\sigma_{{{a}}}^{{b}}q^{R}_{{b}}-
{\textstyle\frac{3}{2}}
[\omega,q^{R}]_{{a}}-{\textstyle\frac{1}{2}}\D^{{b}}\pi^{R}_{{{a}}{{b}}}
 +{\textstyle\frac{1}{3}}\D_{{a}}\mu^{R}
-{\textstyle\frac{1}{3}}\Theta q^{R}_{{a}}\;.
\end{eqnarray}
Gravito-magnetic divergence:
\begin{eqnarray}\label{1+3GrMagDivHO}
 &&\D^{{b}} H_{{{a}}{{b}}}
-(\tilde{\mu}^{m}+\tilde{p}^{m})\omega_{{a}} +[\sigma,E]_{{a}}
 -3E_{{{a}}{{b}}}\omega^{{b}}=-{\textstyle\frac{1}{2}}\curl q^{R}_{{a}}
+(\mu^{R}+p^{R})\omega_{{a}} -{\textstyle\frac{1}{2}}
[\sigma,\pi^{R}]_{{a}} -{\textstyle\frac{1}{2}}\pi^{R}_{{{a}}{{b}}}
\omega^{{b}}\;.
\end{eqnarray}
Standard Matter Conservation (twice contracted Bianchi identities for standard matter)
\begin{eqnarray}\label{cons2}
&&\dot{\mu}^m\,=\, - \,\Theta\,(\mu^m+{p^m})\;,\label{eq:cons1}\\
&&\3nab^{a}{p^m} =  - (\mu^m+{p^m})\,\udot^{a}\,.
\end{eqnarray}
Curvature fluid Conservation (twice contracted Bianchi identities for the curvature fluid )
\begin{eqnarray}
&&\l{eq:cons2} \dot{\mu^R} + \3nab^{a}q^R_{a} = - \,\Th\,(\mu^R+p^R)
- 2\,(\udot^{a}q^R_{a}) -
(\sig^{a}\!^{b}\pi^R_{b}\!_{a})+\mu^{m}\frac{f''\,\dot{R}}{f'^{2}}\;,\\
&& \l{eq:cons3} \dot{q}^R_{\lgl a\rgl} + \3nab_{a}p^R +
\3nab^{b}\pi^R_{ab} = - \,{\textstyle\frac{4}{3}}\,\Th\,q^R_{a} -
\sig_{a}\!^{b}\,q^R_{b} - (\mu^R+p^R)\,\udot_{a} -
\udot^{b}\,\pi^R_{ab} -
\eta_{a}^{bc}\,\om_{b}\,q^R_{c}+\mu^{m}\frac{f''\,\D_{a}{R}}{f'^{2}}\,.
\  \end{eqnarray}

In the equations above the spatial curl of a vector and a tensor is
\begin{equation}
(\c\,X)^{a} = \epsilon^{abc}\,\3nab_{b}X_{c}\,,\qquad \qquad (\c\,X)^{ab} = \epsilon^{cd\lgl a}\,\3nab_{c}X^{b\rgl}\!_{d}\,,
\end{equation}
respectively, where $\epsilon_{abc}=u^d\eta_{abcd}$ is the spatial volume.
Finally $\omega_{{a}}=\frac{1}{2}\ep_{a}{}^{{b}{c}}\omega_{bc}$ and the
covariant tensor commutator is
\[
[W,Z]_{{a}} =\ep_{{{a}}{c}{d}}W^{c}{}_{e} Z^{{d}{e}}\,.
\]

The 1+3 equations above are completely equivalent to the Einstein equation and govern the
dynamics of the matter and gravitational fields in fourth order gravity. As we will see the  new source terms in their R.H.S. will  modify the evolution of the perturbations in a  non-trivial way. The standard GR equations are obtained by setting $f(R)=R$ which corresponds to setting all these sources to zero.
\section{The Covariant Gauge Invariant perturbation equations for scalars}\label{AppB}
In the following we give the equations for the evolution of the gradient variables as done in \cite{SantePertSca, StructForm}
\begin{equation}\label{s3}
{\cal D}^{m}_{{a}}=\frac{S}{\mu^{m}}\D_{{a}}\mu^{m}\,,\qquad
Z_{{a}}=S\D_{{a}}\Theta\,,\qquad C_{{a}}=S^3\D_{{a}}\tilde{R}\;,
\qquad{\cal R}_{{a}}=S\D_{{a}} R\,,\qquad\Re_a=S\D_{{a}} \dot{R}\;,
\end{equation}
They read
\begin{eqnarray}
\dot{{\cal D}}^m_{{a}} &=&w\Theta{\cal D}^m_{{a}}-(1+w)Z_{{a}}\,,\label{s5}\\
\nn\dot{Z}_{{a}} &=& \left(\frac{\dot{R}f''}{f'}-\frac{2 \Theta }{3}\right)Z_a+
   \left[\frac{3 (w -1) (3 w +2)}{6 (w +1)} \frac{\mu^{m}}{ f'} + \frac{2 w \Theta ^2
   +3 w (\mu^{R}+3  p^{R}) }{6 (w +1) }+\frac{2 w}{w +1}\frac{K}{S^2}\right]
   {\cal D}^m_{a}+\frac{\Theta f''}{2f'}\Re_{a}\\&&+
   \left[\frac{1}{2}+2\frac{f''}{f'}\frac{K}{S^2}-\frac{1}{2} \frac{f}{f'}\frac{ f''}{f'}+ \frac{f''}{f'} \frac{\mu^{m}}{
   f'} + \dot{R} \Theta  \left(\frac{f''}{f'}\right)^{2}+ \dot{R} \Theta \frac{ f^{(3)}}{ f'}\right]\mathcal{R}_a
   -\frac{w}{w +1} \3nab^{2}{\cal
   D}^m_{a}-\frac{ f''}{ f'}\3nab^{2}\mathcal{R}_{a}\,,\\
\dot{{\cal R}}_a&=&\Re_{a}-\frac{w }{w +1}\dot{R}\;{\cal D}^m_{{a}}\,,\label{eqZa}\\
\nn\dot{\Re}_a&=&- \left(\Theta + 2\dot{R} \frac{
   f^{(3)}}{f''}\right)\Re_{a}- \dot{R} Z_{a} -
   \left[\frac{ (3 w -1)}{3} \frac{\mu^{m}}{f''} + 3\frac{w}{w +1}
   (p^{R}+\mu^{R}) \frac{f'}{f''}+ \frac{w}{3(w +1)} \dot{R} \left(\Theta
   -3 \dot{R} \frac{f^{(3)}}{f''}\right)\right]{\cal D}^m_{{a}}\\&&\nn+\left[\frac{3}{2} (1+w) \frac{K}{S^2}-\left(
   \frac{1}{3}\frac{f'}{f''}+\frac{f^{(4)}}{f'} \dot{R}^2+\Theta  \frac{f^{(3)}}{f'}
   \dot{R}-\frac{2}{9} \Theta ^2 +\frac{1}{3}(\mu^{R}+3 p^{R}) + \ddot{R} \frac{f^{(3)}}{f''}\right.\right.\\&&\left.\left.
   - \frac{1}{6}\frac{f}{f'}+\frac{1}{2} (w +1) \frac{\mu^{m}}{f'} -\frac{1}{3} \dot{R} \Theta
 \frac{f''}{f'}\right)\right]\mathcal{R}_{a}+\3nab^{2}\mathcal{R}_{a}\,,
\end{eqnarray}
together with the constraint
\begin{equation}\label{Gauss}
  \frac{C_a}{S^2}+ \left(\frac{4  }{3}\Theta +\frac{2 \dot{R}
   f''}{f'}\right) Z_a-2\frac{
    \mu^{m} }{f'}{\cal D}^m_{{a}}+ \left[2 \dot{R}
   \Theta  \frac{f^{(3)}}{ f'}-\frac{f''}{ f'^{2}} \left(f-2 \mu^{m} +2
   \dot{R} \Theta  f''-4\frac{K}{S^{2}}\right)\right]\mathcal{R}_a+\frac{2 \Theta
   f''}{f'}\Re_a-\frac{2 f''}{f'}\3nab^{2}\mathcal{R}_a=0\,.
\end{equation}
The propagation equation for the variable $C_a$ is
\begin{eqnarray}
&& \dot{C}_a=\nonumber \frac{6 K^2 }{S^2 \Theta } \left(5\frac{ f'' }{
   f'}\mathcal{R}_a-3\dd^{m}_a \right)
   +K\left\{\frac{3}{S^2
   \Theta  }C_a +
   \left(\frac{ 6 \mu^{R}}{ \Theta  }-\frac{2 (3\omega)  \Theta }{3(\omega +1)}\right)\dd^m_a-\frac{6  f''}{ \Theta f'}\3nab^{2} \mathcal{R}_a\right.\\&&\left.\nonumber+\left[-\frac{6 R' f''^2}{f'^2}+\frac{\left(2
   \left(\Theta ^2-3 \mu^R\right)
   f'-3 f\right) f''}{\Theta f'^2}+\frac{6 R' f^{(3)}}{f'}\right] \mathcal{R}_a\right\} \\&&+\3nab^{2}\left[\frac{4 \omega
    S^2 \Theta }{3 (\omega +1)}\dd^m_a+\frac{2  S^2
   f''}{f'}\Re_a-\frac{2 S^2 \left(\Theta
    f''-3 \dot{R} f^{(3)}\right)}{3 f'}\mathcal{R}_a\right]\,,
\end{eqnarray}

If we focus on the evolution of scalar part of these variables, which is associated with the spherically symmetric collapse terms and can be extracted taking the divergence of the definitions \rf{s3}, these equations become
\begin{eqnarray}
&&\dot{\Delta}_m =w\Theta \Delta_m-(1+w)Z\,,\label{eqDelta}\\
&&\nn\dot{Z} = \left(\frac{\dot{R}
   f''}{f'}-\frac{2 \Theta }{3}\right)Z+
   \left[\frac{ (w -1) (3 w +2)}{2 (w +1)} \frac{\mu^{m}}{ f'} + \frac{2 w \Theta ^2
   +3 w (\mu^{R}+3  p^{R}) }{6 (w +1) }\right]
   \Delta_m+\frac{\Theta f''}{f'}\Re+\\&&+
   \left[\frac{1}{2}-\frac{1}{2} \frac{f}{f'}\frac{ f''}{f'}- \frac{f''}{f'} \frac{\mu^{m}}{
   f'} + \dot{R} \Theta  \left(\frac{f''}{f'}\right)^{2}+ \dot{R} \Theta \frac{ f^{(3)}}{ f'}\right]\mathcal{R}
   -\frac{w}{w +1} \3nab^{2}{\Delta}_m-\frac{ f''}{f'}\3nab^{2}\mathcal{R}\,,\\
&&\dot{{\cal R}}=\Re-\frac{w }{w +1}\dot{R}\;{\Delta}_m\,,\\
&&\nn\dot{\Re}=- \left(\Theta + 2\dot{R} \frac{
   f^{(3)}}{f''}\right)\Re- \dot{R} Z -
   \left[\frac{ (3 w -1)}{3} \frac{\mu^{m}}{f''} + \frac{w}{3(w +1)} \ddot{R} \right]{\Delta}_m+\\&&-\left[\frac{1}{3}\frac{f'}{f''}+\frac{f^{(4)}}{f'} \dot{R}^2+\Theta \dot{R} \frac{f^{(3)}}{f''}+\ddot{R} \frac{f^{(3)}}{f''}-\frac{R}{3}\right]\mathcal{R}+\3nab^{2}\mathcal{R}\,,\label{eqRho}
\end{eqnarray}
\begin{eqnarray}
&& \dot{C}=\nonumber K^2 \left[\frac{18  f'' \mathcal{R}}{S^2
\Theta f'}-\frac{18\Delta_{m}}{S^2  \Theta } \right] +K\left[\frac{3}{S^2\Theta}C
+\Delta_{m}\left(\frac{2 (w-1) \Theta }{w +1}+\frac{ 6\mu^{R}}{\Theta}\right)-\frac{6  f''}{\Theta f'}\3nab^{2}\mathcal{R}
+\frac{6 f''}{f'}\Re+\right.\\&&
\nonumber\left.+\frac{ 6 \dot{R} \Theta  f'
   f^{(3)}- f'' \left(3 f-2 \left(\Theta ^2-3 \mu^{R}\right) f'+6 \dot{R} \Theta f''\right)}{\Theta  (f')^2 }\mathcal{R}\right] +\3nab^{2}\left[\frac{4 w
    S^2 \Theta }{3 (w +1)}\Delta_{m}+\frac{2  S^2
   f''}{f'}\Re-\frac{2 S^2 \left(\Theta
    f''-3 \dot{R} f^{(3)}\right)}{3
    f'}\mathcal{R}\right]\,,\label{eqC}
\end{eqnarray}
together with the constraint
\begin{equation}\label{Gauss1}
  \frac{C}{S^2}+ \left(\frac{4  }{3}\Theta +\frac{2 \dot{R}
   f''}{f'}\right) Z-2\frac{
    \mu^{m} }{f'}{\Delta_{m}}+ \left[2 \dot{R}
   \Theta  \frac{f^{(3)}}{ f'}-\frac{f''}{ (f')^2} \left(f-2 \mu^{m} +2
   \dot{R} \Theta  f''\right)\right]\mathcal{R}+\frac{2 \Theta
   f''}{f'}\Re-\frac{2 f''}{f'}\3nab^{2}\mathcal{R}=0\,.
\end{equation}
Traditionally the analysis of the perturbation equations is simplified by using a harmonic decomposition.  In the 1+3 formalism this  can be done by developing the scalar quantities defined above using the eigenfunctions of the Laplace-Beltrami operator  \cite{BDE}:
\begin{eqnarray}\label{eq:harmonic}
  \3nab^{2}Q = -\frac{\ell^{2}}{S^{2}}Q\;,
\end{eqnarray}
where $\ell=2\pi S/\lambda$ is the wavenumber and $\dot{Q}=0$. Developing \rf{ScaVar} in terms of $Q$, (\ref{eqDelta}-\ref{Gauss1}) reduce to
\begin{eqnarray}
\dot{\Delta}_{m}^{(\ell)} &=&w\Theta \Delta_{m}^{(\ell)}-(1+w)Z^{(\ell)}\,,\label{eqDeltaHarm}\\
\dot{Z}^{(\ell)} &=& \left(\frac{\dot{R}
   f''}{f'}-\frac{2 \Theta }{3}\right)Z^{(\ell)}+
    \left[\frac{ (w -1) (3 w +2)}{2 (w +1)} \frac{\mu^{m}}{ f'} + \frac{2 w \Theta ^2
   +3 w (\mu^{R}+3  p^{R}) }{6 (w +1) }\right]   \Delta_{m}^{(\ell)}+\frac{\Theta f''}{f'}\Re^{(\ell)}+\nonumber \\&&+
   \left[\frac{1}{2}-\frac{ f''}{f'} \frac{\ell^2}{S^2}-\frac{1}{2} \frac{f}{f'}\frac{ f''}{f'}- \frac{f''}{f'} \frac{\mu^{m}}{
   f'} + \dot{R} \Theta  \left(\frac{f''}{f'}\right)^{2}+ \dot{R} \Theta \frac{ f^{(3)}}{ f'}\right]\mathcal{R}^{(\ell)}\,,\\
\dot{{\cal R}}^{(\ell)}&=&\Re^{(\ell)}-\frac{w }{w +1}\dot{R}\;{\Delta}_{m}^{(\ell)}\,,\label{eqZHarm}\\
\dot{\Re}^{(\ell)}&=&- \left(\Theta + 2\dot{R} \frac{f^{(3)}}{f''}\right)\Re^{(\ell)}- \dot{R} Z^{(\ell)} -
  \left[\frac{ (3 w -1)}{3} \frac{\mu^{m}}{f''} + \frac{w}{3(w +1)} \ddot{R} \right]{\Delta}_{m}^{(\ell)}+\nonumber\\ &&+\left[\frac{\ell^{2}}{S^2}-\left(\frac{1}{3}\frac{f'}{f''}+\frac{f^{(4)}}{f'} \dot{R}^2+\Theta \dot{R} \frac{f^{(3)}}{f''}+\ddot{R} \frac{f^{(3)}}{f''}-\frac{R}{3}\right)\right]\mathcal{R}^{(\ell)}\,,\label{eqRho2}\\
\dot{C}^{(\ell)}&=&\nn K^2 \left[\frac{18  f'' \mathcal{R}}{S^2
\Theta f'}-\frac{18\Delta_{m}}{S^2  \Theta } -6\frac{f''}{\Theta f'}\Re^{(\ell)}\right]  +K\left[\frac{3}{S^2\Theta}C^{(\ell)}
+\Delta^{(\ell)}\left(\frac{2 (w-1) \Theta }{w +1}-\frac{ 6\mu^{R}}{\Theta}\right)+\frac{6 f''}{f'}\Re^{(\ell)} +\right.\\&&\nonumber\left.+\frac{ 6 S^{2} \dot{R} \Theta  f' f^{(3)}-6 \ell^2 f'' f'+  S^{2} f'' \left(3 f-2 \left(\Theta ^2-3 \mu^{R}\right) f'+6 \dot{R} \Theta f''\right)}{\Theta S^{2} (f')^2 }\mathcal{R}^{(\ell)}\right]+\nn\\&&+\frac{\ell^{2}}{S^{2}}\left[\frac{4 w S^2 \Theta }{3 (w +1)}\Delta_{m}^{(\ell)}+\frac{2  S^2
   f''}{f'}\Re^{(\ell)}-\frac{2 S^2 \left(\Theta f''-3 \dot{R} f^{(3)}\right)}{3 f'}\mathcal{R}^{(\ell)}\right]\,,\label{eqCHarm}
    \end{eqnarray}
\begin{eqnarray}
 0&=& \frac{C^{(\ell)}}{S^2}+ \left(\frac{4  }{3}\Theta +\frac{2
\dot{R}
   f''}{f'}\right) Z^{(\ell)}-2\frac{
    \mu^{m} }{f'}{\Delta}_m^{(\ell)}+ \left[2 \dot{R}
   \Theta  \frac{f^{(3)}}{ f'}-\frac{f''}{ (f')^2} \left(f-2 \mu^{m} +2
   \dot{R} \Theta  f''\right)+2 \frac{ f''}{f'} \frac{\ell^{2}}{S^2}\right]\mathcal{R}^{(\ell)}+\frac{2 \Theta
   f''}{f'}\Re^{(\ell)}\;,\nn\\\label{constrCZDel}
\end{eqnarray}
which is a system of ordinary differential equations. This system takes a more manageable form if we reduce it to a pair of second order equations:
\begin{eqnarray}
   &&\nn\ddot{\Delta}_{m}^{(\ell)}+\left[\left(
   \frac{2}{3}-w\right) \Theta -\frac{\dot{R}
   f''}{f'}\right] \dot{\Delta}_{m}^{(\ell)}-\left[w  \frac{\ell^2}{S^{2}}-w  (3
   p^{R}+\mu^{R})-\frac{2 w  \dot{R} \Theta
   f''}{f'}-\frac{\left(3 w ^2-1\right) \mu^{m} }{f'}\right]\Delta_{m}^{(\ell)}=\\&&=
   \frac{1}{2}(w +1)\left[-2 \frac{\ell^2}{S^2}\frac{f''}{f'}-1+
   \left(f-2 \mu^{m} +2 \dot{R} \Theta  f''\right)\frac{f''}{f'^2}
   -2  \dot{R} \Theta
   \frac{f^{(3)}}{f'}\right] \mathcal{R}^{(\ell)} -\frac{(w +1) \Theta
   f'' }{f'}\dot{\mathcal{R}}^{(\ell)}\label{EqPerIIOrd1}\,,\\&&
   \nn f''\ddot{\mathcal{R}}^{(\ell)}+\left(\Theta f'' +2 \dot{R}
   f^{(3)}\right)
   \dot{\mathcal{R}}^{(\ell)}-\left[\frac{\ell^2}{S^2}f''+ 2 \frac{K}{S^2}f''
  +\frac{2}{9} \Theta^2 f''- (w +1) \frac{\mu^{m}}{2 f'}f''- \frac{1}{6}(\mu^{R}+ 3
p^{R})f''+\right.\\&&\nn\left.-\frac{f'}{3}+ \frac{f}{6 f'}f'' +
\dot{R} \Theta  \frac{f''^{2}}{ f'} -
   \ddot{R} f^{(3)}- \Theta f^{(3)} \dot{R}- f^{(4)}\dot{R}^2
   \right]\mathcal{R}^{(\ell)}=-
   \left[ \frac{1}{3}(3 w -1) \mu^{m}+ \right.\\&&\left.+\frac{w}{1+w} \left(f^{(3)}
   \dot{R}^2+  (p^{R}+\mu^{R}) f'+ \frac{7}{3}\dot{R} \Theta f''
 +\ddot{R} f'' \right)\right]\Delta_{m}^{(\ell)}-\frac{(w -1) \dot{R} f''}{w +1} \dot{\Delta}_{m}^{(\ell)}\,.\label{EqPerIIOrd2}
\end{eqnarray}

\section{The Covariant Gauge Invariant perturbation equations for tensors}\label{AppC}
These equations were derived in  \cite{K1} here we sketch briefly their derivation.
If one focus only on tensor perturbations the 1+3 equations reduce to
\begin{equation}\label{eqsigma}
    \dot{\sigma} _{ab} + \frac{2}{3}\,\Theta \,\sigma _{ab}
  + E_{ab} - \frac{1}{2} {\pi}_{ab}=0\;,
\end{equation}
\begin{equation}\label{eqmag}
   \dot{H}_{ab}+
   H_{ab}\,\Theta+(\c\,E)_{ab} - \frac{1}{2}(\c\,{\pi})_{ab} = 0\;,
\end{equation}
\begin{equation}\label{eqelect}
   \dot{E}_{ab}+E_{ab}\,\Theta -
  (\c\,H)_{ab}  +
   \frac{1}{2}\left({\mu^{tot}} +{p^{tot}} \right) \,\sigma_{ab}
   + \frac{1}{6}\Theta\,{\pi}_{ab} +
   \frac{1}{2} \dot{{\pi}}_{ab}=0\;,
\end{equation}
together with the conditions
\begin{equation}\label{constraints}
\3nab_{b}H^{ab}=0\;,\quad \3nab_{b}E^{ab}=0\;,\quad
H_{ab}=(\c\,\sigma)_{ab}\;.
\end{equation}
Taking the time derivative of the above
equations we obtain
\begin{equation}\label{eq2ordSigma}
\ddot{\sigma}_{ab}-\3nab^{2}\sigma  + \frac{5}{3}\, \Theta \,
\dot{\sigma}_{ ab} +  \left(\frac{1}{9}\, {\Theta }^2 +
\frac{1}{6}{\mu^{tot}} - \frac{3}{2}{p^{tot}}  \right)\, \sigma_
{ab}=\dot{{\pi}}_{ab}+ \frac{2}{3}\, \Theta \, {\pi}_{ab}\;,
\end{equation}
\begin{equation}\label{eq2ordMag}
\ddot{H}_{ab} - \3nab^{2}H_{ab} + \frac{7}{3}\, \Theta\,
\dot{H}_{ab} + \frac{2}{3}\, \left({\Theta }^2 - 3 {p}^{tot} \right)\,
H_{ab} = (\c\, \dot{{\pi}})_{ab}+\frac{2 }{3}\, \Theta\, (\c\, \
{\pi})_{ab}\;,
\end{equation}
\begin{eqnarray}\label{eq2ordElect}
\nn \ddot{E}_{ab}&-&\3nab^{2}E_{ab}+
\frac{7}{3}\,\Theta\,\dot{E}_{ab} +\frac{2}{3}\, \left( {\Theta}^2 -
3 {p} \right) {E}_{ab}  -  \frac{1 }{6}\Theta\,\left({\mu^{tot}}
+ {p}^{tot} \right) \,\left( 1 + 3\,c_{s}^{2}\right)\,\sigma _{ab} \\
&=& -
\left[\frac{1}{2}\ddot{{\pi}}_{ab}-\frac{1}{2}\3nab^{2}{\pi}_{ab}
+\frac{5}{6}\,\Theta\,\dot{{\pi}}_{ab} + \frac{1}{3}\left(
{\Theta }^2 - {\mu^{tot}} \right) \,{\pi}_{ab}\right]\;,
\end{eqnarray}
where the sound speed  is defined as $c_{s}^{2}=\dot{p}^{tot}/\dot{\mu}^{tot}$ and we have used the
Raychaudhuri equation (Eq.~(\ref{1+3eqRayHO})), the matter conservation
equation (Eq.~(\ref{cons2})) and the commutator identity
\begin{equation}\label{}
(\c\,\dot{X})_{ab}=(\c\,X)^{\cdot}_{ab}+
  \frac{1}{3}(\c\,X)\,\Theta\;.
\end{equation}
Using trace-free symmetric
tensor eigenfunctions of the spatial  the Laplace-Beltrami operator
defined by:
\begin{eqnarray}
\3nab^{2}Q_{ab} = -\frac{\ell^{2}}{a^{2}}Q_{ab}\;,
\end{eqnarray}
where $\ell=2\pi S/\lambda$ is the wavenumber and $\dot{Q}_{ab}=0$ one obtains
\begin{eqnarray}\label{SigmaEqF}
\ddot{\sigma}^{(\ell)} + \left( \frac{5}{3}{\Theta}
+\dot{R}\frac{f''}{f'} \right)\dot{\sigma}^{(\ell)} + &&\left\{
\frac{1}{9}\, {{\Theta} }^2 + \frac{1}{f'}\left(\frac{1}{6}\mu^{m}
-\frac{3}{2}p^{m}\right) +\frac{\ell^2}{a^2} \right.\nonumber\\
&& \left.- \frac{1}{2}{\Theta}\dot{R}\frac{f''}{f'}
-\frac{5}{6}\frac{1}{f'}\left(f-f'R\right) -\dot{R}^{2} \left[
\frac{1}{2}\frac{f'''}{f'} + \left(\frac{f''}{f'}\right)^2 \right]
-\frac{1}{2}\ddot{R}\frac{f''}{f'} \right\} \sigma^{(\ell)}=0,
\end{eqnarray}
\begin{eqnarray}\label{HEqF}
\ddot{H}^{(\ell)} + \left( \frac{7}{3}{\Theta} +\dot{R}\frac{f''}{f'}
\right)\dot{H}^{(\ell)} +  &&\left\{ \frac{2}{3}\, {{\Theta} }^2
-\frac{2}{f'}p^{m} +\frac{\ell^2}{a^2} -\frac{1}{3}
{\Theta}\dot{R}\frac{f''}{f'} \right.\nonumber\\
&& \left. -\frac{1}{f'}\left(f-f'R\right) - \dot{R}^{2} \left[
\frac{f'''}{f'} + \left(\frac{f''}{f'}\right)^2 \right]
-\ddot{R}\frac{f''}{f'} \right\} H^{(\ell)}  = 0,
\end{eqnarray}
\begin{eqnarray}\label{EEqF}
&& {E}^{(\ell)} = -\dot{\sigma}^{(\ell)} - \left(\frac{2}{3}\Theta +
\frac{1}{2}\dot{R}\frac{f''}{f'} \right)\sigma^{(\ell)}.
\end{eqnarray}
where we have used \rf{eqsigma}  as an equation for $E_{ab}$  and we have used the fact
that the tensor part of the anisotropic pressure is $\pi_{ab}=\frac{f''}{f'} \sigma_{a
b}\dot{R}$. It is clear that the first two equations are closed and the third is only a constraint. Therefore the only relevant equation in our case is \rf{SigmaEqF}.

\end{document}